\documentclass[useAMS,usenatbib]{mn2e}
\usepackage{epsfig}
\usepackage{graphicx}
\usepackage{times}
\usepackage{color}
 \usepackage{hhline}

\def\cm3{cm$^{-3}$}
\def\kms{km~s$^{-1}$}

\def\msun{M$_{\odot}$}

\def\one{{\,\sc i}}
\def\two{{\,\sc ii}}
\def\three{{\,\sc iii}}

\def\beq{\begin{equation}}
\def\eeq{\end{equation}}

\def\lesssim{\mathrel{\hbox{\rlap{\hbox{\lower4pt\hbox{$\sim$}}}\hbox{$<$}}}}
\def\gtrsim{\mathrel{\hbox{\rlap{\hbox{\lower4pt\hbox{$\sim$}}}\hbox{$>$}}}}

\def\cmfgen{{\sc cmfgen}}

\def\v1d{{\sc v1d}}

\newcommand{\iso}[2]{\ensuremath{^{#1}\rm{#2}}}

\def\aj{AJ}
\def\pasp{PASP}
\def\apj{ApJ}
\def\apjs{ApJS}
\def\apjl{ApJL}
\def\aap{A\&A}

\def\mnras{MNRAS}
\def\nat{Nature}

\title[Faint and fast-decaying SNe]{One-dimensional non-LTE time-dependent radiative transfer of an He-detonation model
and the connection to faint and fast-decaying supernovae}
\author[Luc Dessart and D. John Hillier]{Luc Dessart$^{1}$ and D. John Hillier$^{2}$ \\ \\
$^1$: Laboratoire Lagrange, UMR\,7293, Universit\'e Nice Sophia-Antipolis, CNRS,
Observatoire de la C\^{o}te dÕAzur, 06300 Nice, France. \\
$^2$: Department of Physics and Astronomy \& Pittsburgh Particle physics, Astrophysics, and Cosmology Center
(PITT PACC), \\
University of Pittsburgh, 3941 OÕHara Street, Pittsburgh, PA 15260, USA
}

\date{Accepted 2014 November 25.  Received 2014 November 24; in original form 2014 October 13}

\pagerange{\pageref{firstpage}--\pageref{lastpage}} \pubyear{2014}

\begin{document}

\maketitle

\label{firstpage}

\begin{abstract}
  We present non-LTE time-dependent  radiative transfer simulations for ejecta produced by the detonation 
  of an helium shell
  at the surface of a low-mass carbon/oxygen white dwarf (WD). This mechanism is one possible origin for
  supernovae (SNe) with faint and fast-decaying light curves, such as  .Ia SNe and Ca-rich transients.
  Our initial ejecta conditions at 1\,d are given by the 0.18\,B explosion model COp45HEp2 of Waldman et al..
  The  0.2\,\msun\ ejecta initially contains 0.11\,\msun\ of He, 0.03\,\msun\ of Ca, and 0.03\,\msun\ of Ti.
  We obtain a $\sim$5\,d rise to a bolometric maximum of 3.59$\times$10$^{41}$\,erg\,s$^{-1}$, primarily
  powered by \iso{48}V decay. Multi-band light curves show distinct morphologies, with a rise to maximum
  magnitude ($-$14.3 to $-$16.7\,mag) that varies between 3 to 9\,d  from the $U$ to the $K$ bands. Near-IR light curves
  show no secondary maximum. 
  Because of the presence of both He\one\ and Si\two\ lines at early times we obtain a hybrid Type Ia/Ib classification.
  During the photospheric phase line blanketing is caused primarily by Ti\two. At nebular times,
  the spectra show strong Ca\two\ lines in the optical (but no [O\one] 6300--6364\,\AA\ emission), and Ti\two\ in the near-IR.
  Overall, these results match qualitatively the very disparate properties of .Ia SNe and Ca-rich transients.
  Although the strong Ti\two\ blanketing and red colors that we predict are rarely observed, they are seen, for example, 
  in OGLE-2013- SN-079.
  Furthermore, we obtain a faster light-curve evolution than, for example, PTF10iuv, indicating an ejecta 
  mass $>$\,0.2\,\msun.
  An alternate scenario may be the merger of two WDs,  one or both composed of He.
 \end{abstract}


\section{Introduction}

Helium-shell detonations at the surface of CO white dwarfs (WDs) have been
studied a number of times over the last 30 years (see, e.g.,
\citealt{taam_80,woosley_weaver_94,livne_glasner_90,livne_glasner_91,livne_arnett_95,bildsten_etal_07}).
The ignition conditions are complex, making the subsequent evolution of the burning shell
uncertain. Further, the properties of the accreted helium shell at detonation depends on the CO WD mass.
These variations in properties are known to considerably alter the nucleosynthetic yields, with the
production of intermediate-mass elements (IMEs) favored at lower density (see, e.g., \citealt{shen_etal_10,waldman_etal_11})
and iron-group elements (IGEs) favored at higher densities \citep{fink_etal_07}.
Furthermore, for higher mass CO WDs, the surface detonation may cause the detonation of the
core, leading to a full disruption of the system rather than a shell ejection.
Lacking a well defined model for the detonations of such shells
(e.g., initial density/radial structure for the shell, mechanism of ignition, evolution of the combustion, multi-dimensional effects, etc. ),
it is unclear what scenarios occur, and with what frequencies. Although He-shell detonations are unlikely
to produce the population of standard SNe Ia \citep{fink_etal_07}, they represent
physical conditions that may occur in Nature and thus warrant study.

Renewed interest in helium-shell detonations has come from the recent discovery of rare transients characterized by
\begin{enumerate}
\item a fast-evolving light curve with a modest peak brightness (by SN standards);
\item a peculiar composition revealed through the presence of spectral signatures from helium and IMEs (e.g., Ti, Ca, Sc)
and inferred (for SN\,2005E) from nebular line analyses \citep{perets_etal_10,sullivan_etal_11,kasliwal_etal_12};
\item the Ca\two-dominated optical spectra at nebular times, with no signature from IGEs.
\end{enumerate}
They exhibit standard line widths during the high-brightness phase,
 suggesting the expansion rate is comparable to standard-energy SNe.
 The faster evolving light curves suggest, however, a  lower ejecta mass.
These extraordinary SNe Ib are associated with old stellar populations
\citep{perets_etal_11b} located at large distances from their hosts \citep{kasliwal_etal_12},
rather than with star forming regions rich in massive stars \citep{anderson_james_09}.
Prototypical events of that class are SN\,2005E, SN\,2007ke, PTF\,09dav, PTF\,10iuv, and PTF\,11bij.

Another class of fast-evolving transients are Type Iax SNe --- the prototypical member of the SN Iax class is SN\,2002cx.
Relative to standard SNe Ia, these SNe have narrower spectral lines and are fainter by $\gtrsim$\,1\,mag at maximum;
they also have blue early-time spectra reminiscent of SN\,1991T \citep{foley_etal_13}.
The origin of SNe Iax is debated. They may stem from low-energy deflagrations \citep{jordan_etal_12,fink_etal_14},
or perhaps double detonations in sub-Chandrasekhar white dwarfs (see, e.g., \citealt{sim_etal_12}).
Their spectra differ from SN\,2005E, with the
most extreme narrow lined member, SN\,2008ha, having a spectrum dominated by IMEs.
\citet{foley_etal_13} excluded SN\,2005E from the Iax class on the basis of its somewhat distinct spectra from
the prototype SN\,2002cx and
its association with an early-type galaxy since Type Iax SNe tend to be associated with late-type galaxies.
\citet{foley_etal_13} raise the possibility that SN\,2005E arose through accretion from a degenerate He star, rather
than from a non-degenerate He star proposed for the Iax class.

Radiative-transfer simulations of helium-shell detonations have been performed
by \citet{perets_etal_10}, who focused on nebular-phase spectra to emphasize the large
abundance of Ca over IGEs and the unique nature of these events.
\citet{waldman_etal_11} performed a set of simulations for 0.15-0.3\,\msun\, He shells
detonating on low-mass CO WDs of 0.4-0.6\,\msun. Their radiative-transfer modeling is centered on their
model COp45HEp2 and a confrontation to SN\,2005E. \citet{shen_etal_10} performed a set of simulations for similar
He shells but typically associated with higher mass CO WDs, leading to large differences in yields, light curves,
and spectra. Despite the diversity of ejecta and radiation properties (and the mismatches to observations),
these simulations suggest that He-shell detonations provide a promising framework for understanding
.Ia SNe, Ca-rich transients, and more generally faint and fast transients associated with white-dwarf explosions.

In this paper we revisit these simulations. We focus on model COp45HEp2 of \citet{waldman_etal_11}
but this time perform non-Local-Thermodynamic-Equilibrium (non-LTE)
time-dependent simulations of the full ejecta. This multi-epoch study extends
from 1.1 until $\sim$\,60\,d after the explosion. The non-LTE approach takes explicit account of non-thermal processes
and allows for $\gamma$-ray escape and non-local energy deposition. This is critical since helium represents
about one half of the ejecta mass and this new class of fast transients shows He\one\ lines 
(hence the SN Ib classification) ---
all simulations so far have ignored non-thermal processes and produced synthetic spectra with no He\one\ lines.
Model bolometric and multi-band photometric light curves, and spectra, are compared to the well observed
SN PTF\,10iuv \citep{kasliwal_etal_12}. To address the spectral diversity of this class, we also compare 
the model to SNe\,2005E \citep{perets_etal_10} and PTF\,09dav \citep{sullivan_etal_11}.

In Section~\ref{sect_obs}, we present the observational data we use in this paper.
The numerical setup and the initial ejecta
conditions for our radiative transfer simulations with \cmfgen\ are discussed in Section~\ref{sect_setup}.
For details on our approach, we refer the reader to
\citet{DH05_qs_SN,DH08_time,DH10,DH11,dessart_etal_12,li_etal_12,HD12,D14}.
We then present the results from our simulations, first describing the evolution of the ejecta properties
(Section~\ref{sect_ejecta}), and then those of the radiation, for both photometry (Section~\ref{sect_phot})
and spectroscopy (Section~\ref{sect_spec}).
We digress on the estimate of the calcium mass and the oxygen mass from nebular spectra in
Sections~\ref{sect_ca} and \ref{sect_oxy}.
In Section~\ref{sect_comp}, we confront our model results to observations and present our conclusions.

\begin{table*}
\caption{Summary of ejecta properties for the helium detonation model COp45HEp2.
We give the masses for the different species at a post-explosion time of 1.16\,d, which is the start time
of our \cmfgen\ simulations.
However, for the isotopes that are unstable (last four columns; subscript o), we give their mass immediately after the
combustion stops, i.e., prior to any decay. This is why the \iso{48}Cr$_{\rm o}$ (given at 0\,d) is larger than the Cr mass (given at 1.16\,d).
\label{table_comp_init}
}
\begin{tabular}{c@{\hspace{3mm}}c@{\hspace{3mm}}c@{\hspace{3mm}}c@{\hspace{3mm}}c@{\hspace{3mm}}c@{\hspace{3mm}}
c@{\hspace{3mm}}c@{\hspace{3mm}}c@{\hspace{3mm}}c@{\hspace{3mm}}c@{\hspace{3mm}}c@{\hspace{3mm}}c@{\hspace{3mm}}
c@{\hspace{3mm}}}
\hline
E$_{\rm kin}$  &   M$_{\rm ejecta}$  & He &     C &   O &    Si &     Ca &   Sc &   Ti &      Cr &    \iso{44}Ti$_{\rm o}$ &   \iso{48}Cr$_{\rm o}$
&   \iso{52}Fe$_{\rm o}$ &  \iso{56}Ni$_{\rm o}$ \\
 ${\rm [B]}$  &     [\msun] &  [\msun] &  [\msun] & [\msun] & [\msun] & [\msun] & [\msun] & [\msun] & [\msun] & [\msun] & [\msun] & [\msun] & [\msun] \\
\hline
0.18 &           0.2 &      1.12(-1) &   5.46(-4) &  2.32(-6) & 1.03(-3) &  3.39(-2) & 1.71(-5) &  3.29(-2) &  2.68(-3) &  3.25(-2) &  5.60(-3) &  8.64(-4) &  1.13(-4)   \\
 \hline
 \end{tabular}
 \end{table*}

\begin{figure}
\epsfig{file=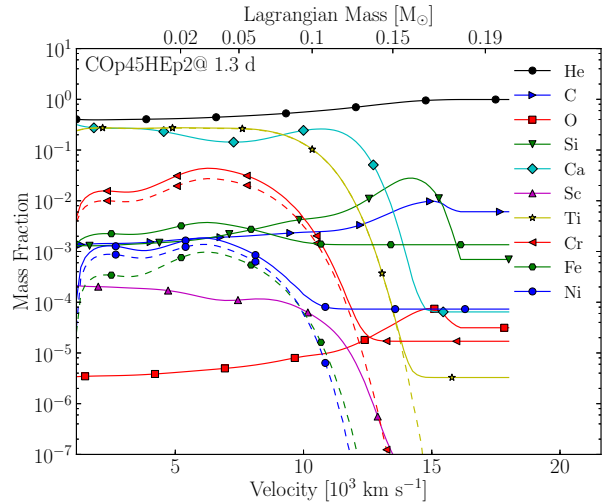,width=9cm}
\caption{Composition of the ejecta model COp45HEp2 at a time of 1.16\,d after explosion as a
function of velocity (bottom axis) and Lagrangian mass coordinate  (top axis).
We also show the contributions of the dominant unstable isotope (dashed lines) for species Ti, Cr, Fe, and Ni.
These isotopes are \iso{44}Ti (yellow/star), \iso{48}Cr (red/triangle), \iso{52}Fe (green/hexagon), and \iso{56}Ni
(blue/circle).
\label{fig_comp_init}
}
\end{figure}

 \begin{figure*}
\epsfig{file=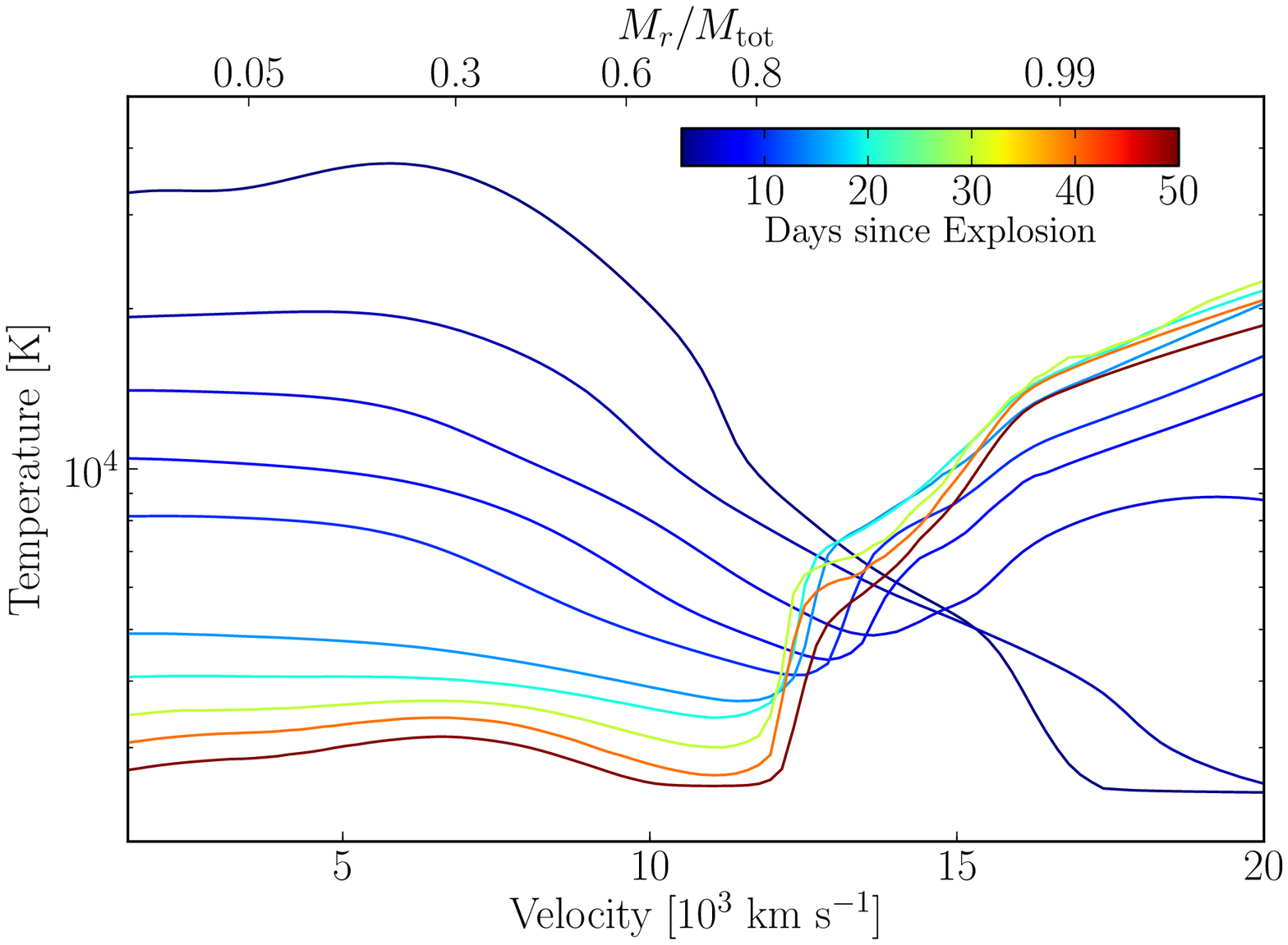,width=8.5cm}
\epsfig{file=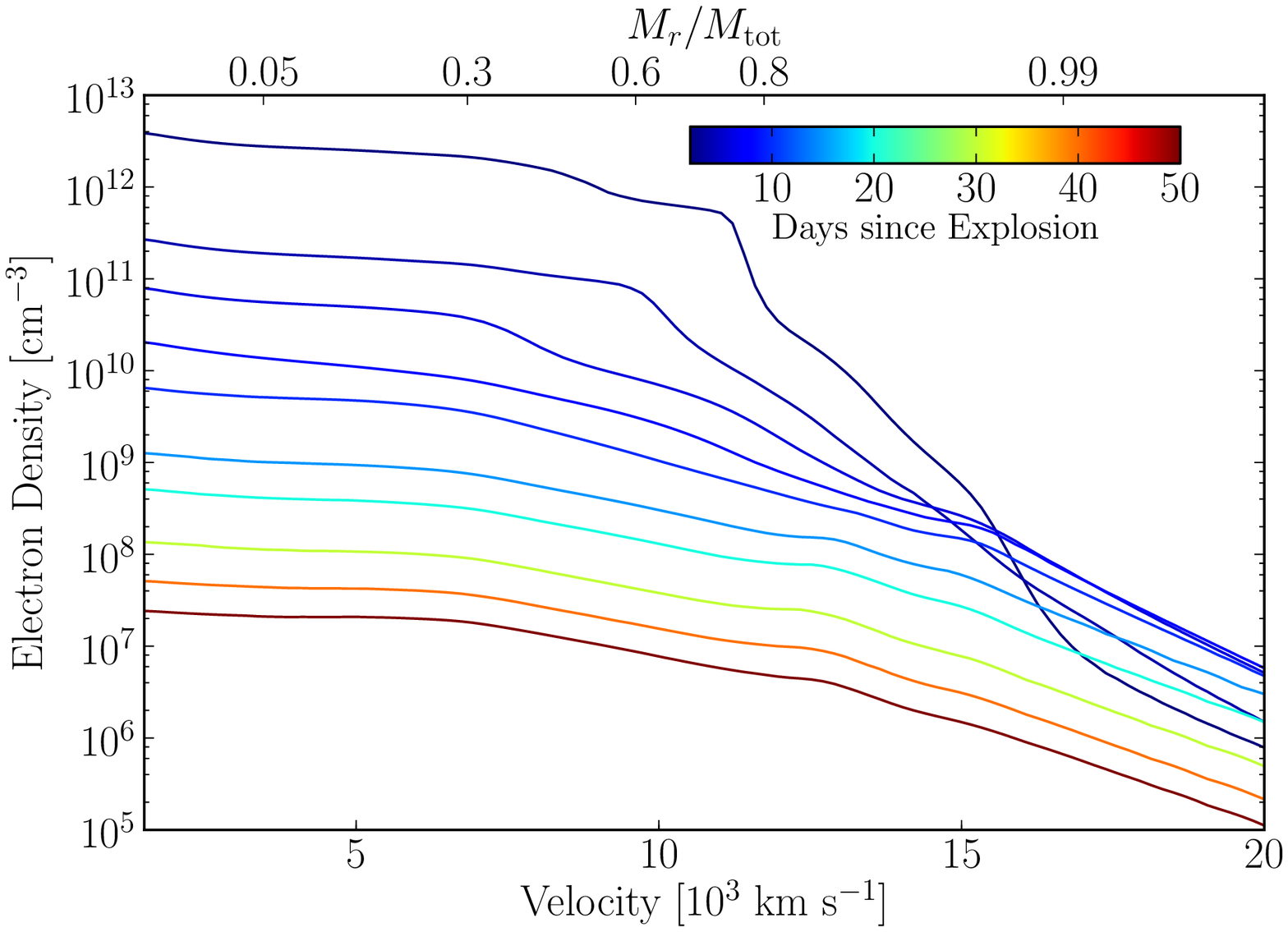,width=8.5cm}
\epsfig{file=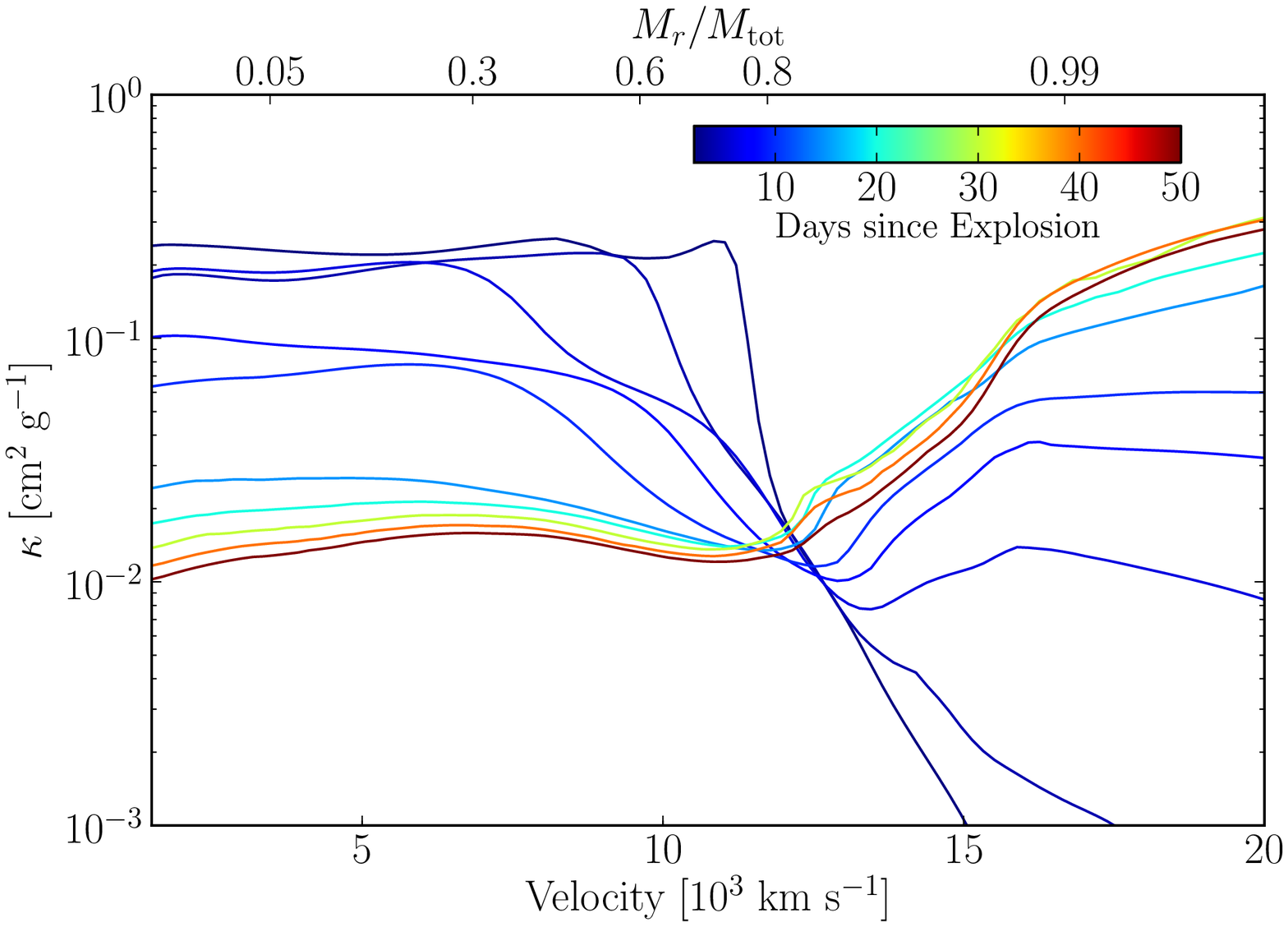,width=8.5cm}
\epsfig{file=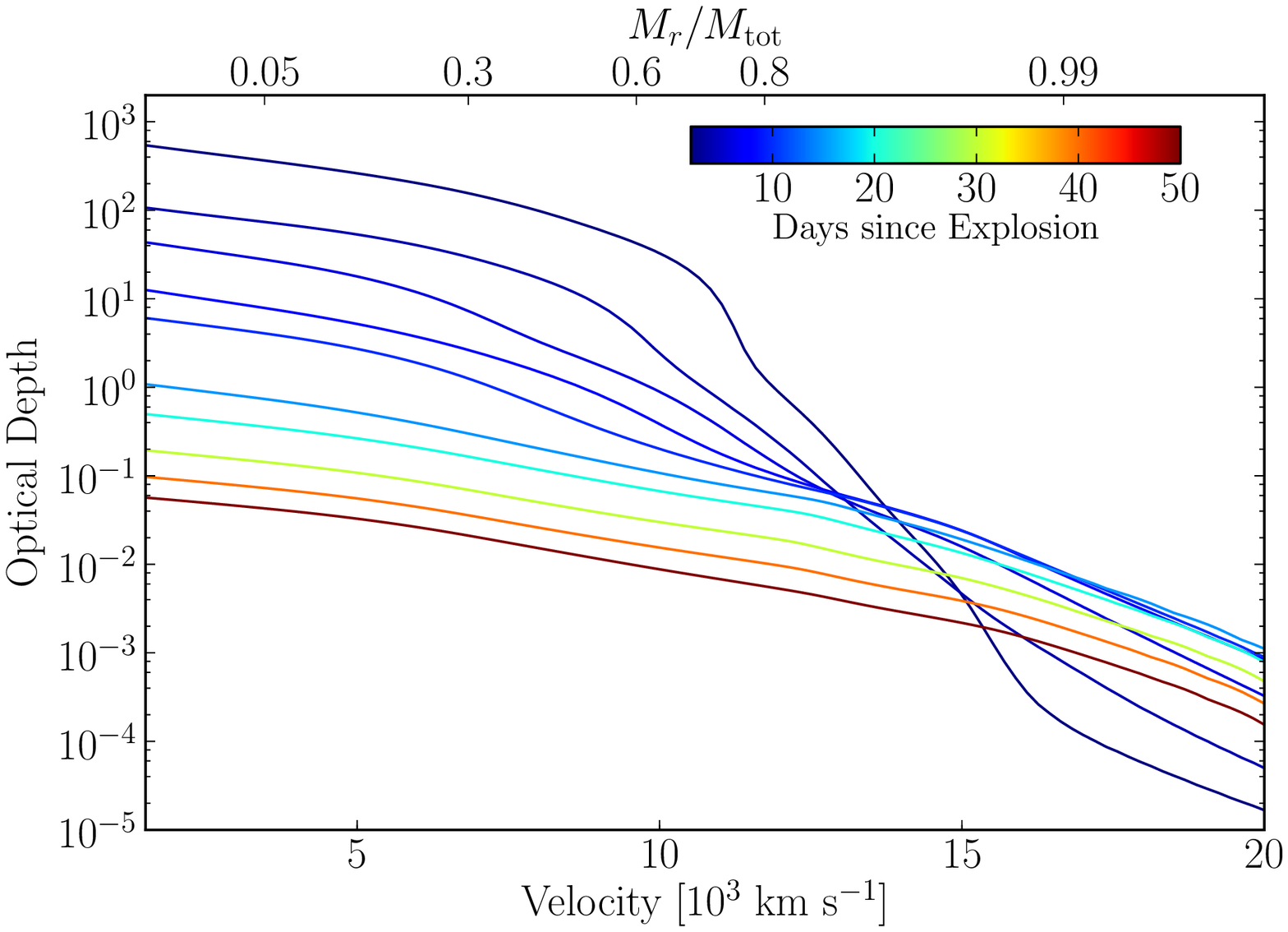,width=8.5cm}
\caption{Evolution versus velocity (bottom axis) and fractional Lagrangian mass (top axis)
of the ejecta temperature, the electron density,
the Rosseland-mean opacity $\kappa$ and the associated optical depth.
The times shown are 2, 4, 6, 8, 10, 15, 20, 30, 40, 50\,d after explosion.
The total ejecta mass $M_{\rm tot}$ is 0.2\,\msun.
\label{fig_prop_evol}
}
\end{figure*}

\section{Observations}
\label{sect_obs}

  To confront our model predictions to observations, we select a few fast evolving transients,
 including PTF10iuv \citep{kasliwal_etal_12}, SN\,2005E \citep{perets_etal_10}, and
 PTF09dav \citep{sullivan_etal_11}.
 For PTF10uiv, we adopt a distance modulus of 35.1, a redshift of 0.0251485, and we assume
 no reddening \citep{kasliwal_etal_12}.
 For SN\,2005E (PTF\,09dav), we use a redshift of 0.00816565 (0.04) and neglect reddening
 (which is believed to be very small; \citealt{perets_etal_10,sullivan_etal_11}).
  We also adopt a distance of 34\,Mpc for SN\,2005E.  The data shown in this paper was
  retrieved from the  {\sc wiserep} website \citep{wiserep}.

\section{Numerical setup and initial conditions}
\label{sect_setup}

  \citet{waldman_etal_11} studied the detonation of helium shells with a mass in the range 0.15-0.3\,\msun\
resting at the surface of CO WDs of 0.4-0.6\,\msun. Based on the rough agreement of their model COp45HEp2 model
with SN\,2005E, considered as the prototype of Ca-rich transients \citep{perets_etal_10}, we utilize the same model.
In the future, we will investigate other configurations
to characterize common and distinct properties of the members of that class.

  The ejecta model COp45HEp2 is characterized by a mass of 0.2\,\msun\ and a kinetic energy of 0.18\,B.
 The total yields for the main species are given in Table~\ref{table_comp_init}. Helium represents half the
total ejecta mass while carbon has a mass fraction of $\sim$\,0.001 throughout the ejecta (a value close to solar).
Oxygen is not a final yield of the combustion and ends up with an underproduction of $\sim$\,100
compared to solar.
For IMEs, the situation is disparate with elements such as Si and S having mass fractions on the
order of only $\sim$\,0.001 (thus significantly less than for the combustion of CO-rich material in SNe Ia
progenitors but nonetheless somewhat larger than the solar values).
On the other hand, for Ti, Ca, and Cr, the overproduction compared to solar metallicity, prior to any decay, is on the
order of 10$^3$-10$^4$.
This contrast is easily seen in Fig.~\ref{fig_comp_init} when comparing the composition between the deep
layers of the ejecta (affected by combustion) and the outermost layers (where we adopt a solar mixture for metals).

 \begin{figure*}
\epsfig{file=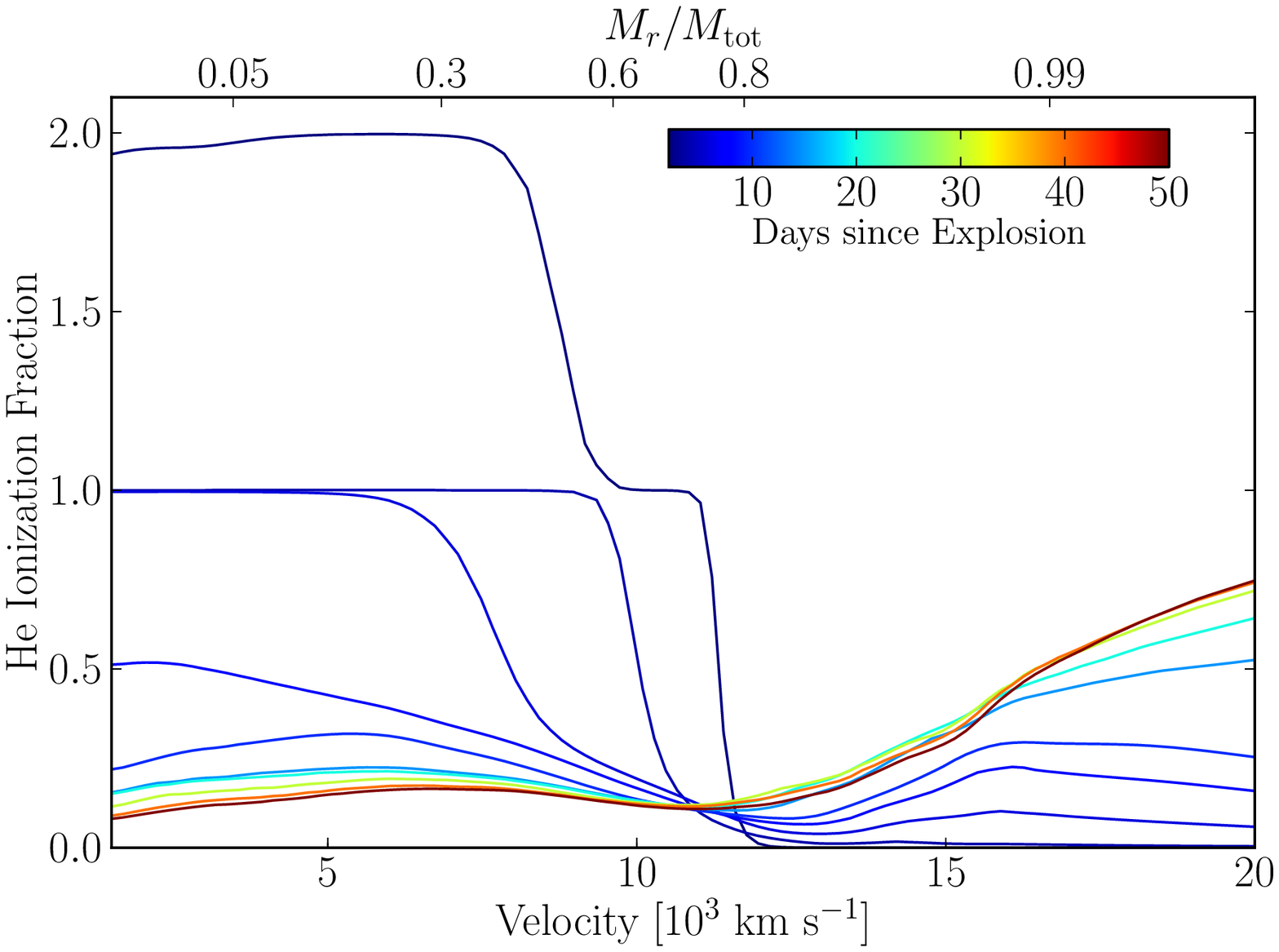,width=8.5cm}
\epsfig{file=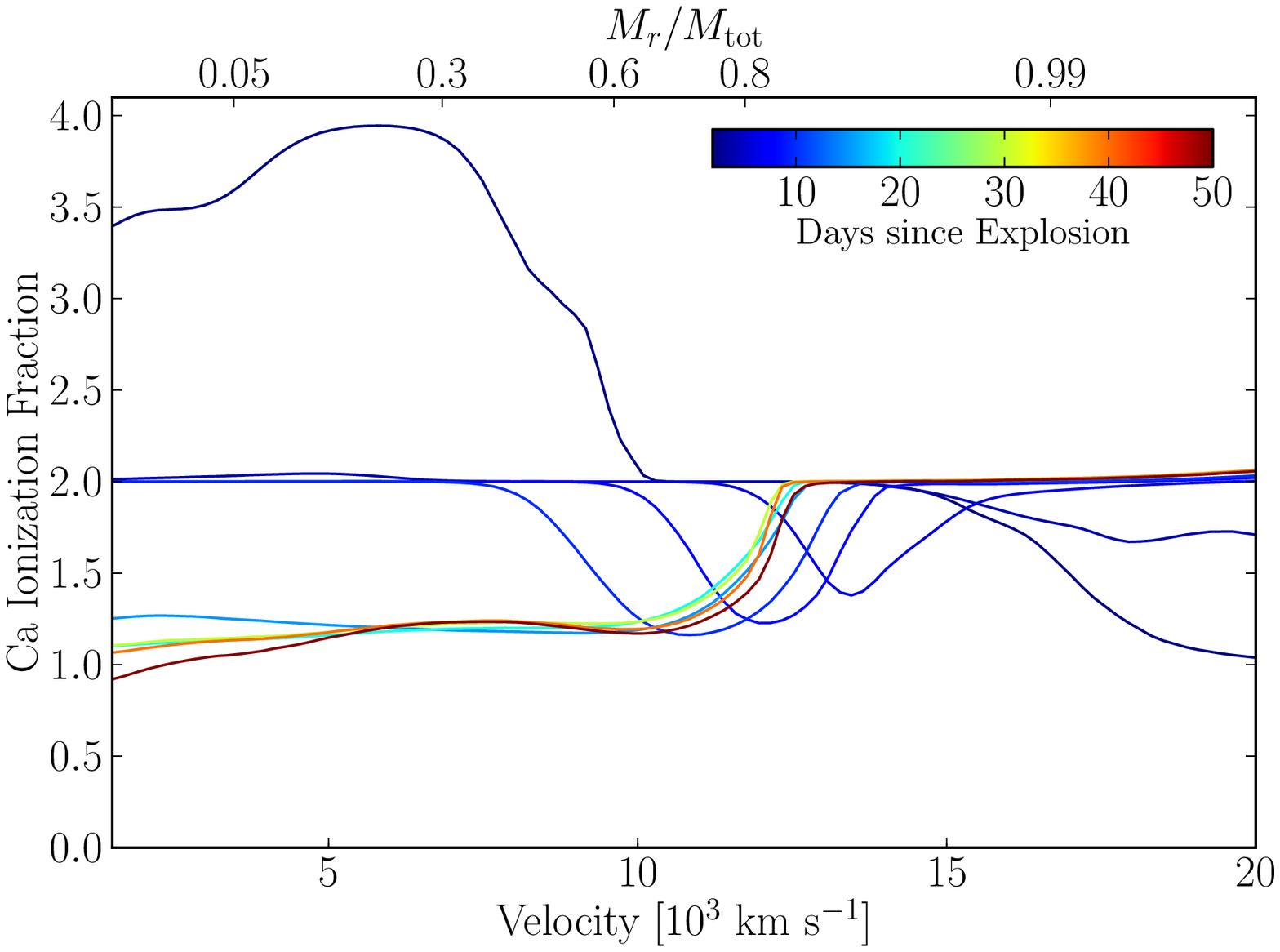,width=8.5cm}
\epsfig{file=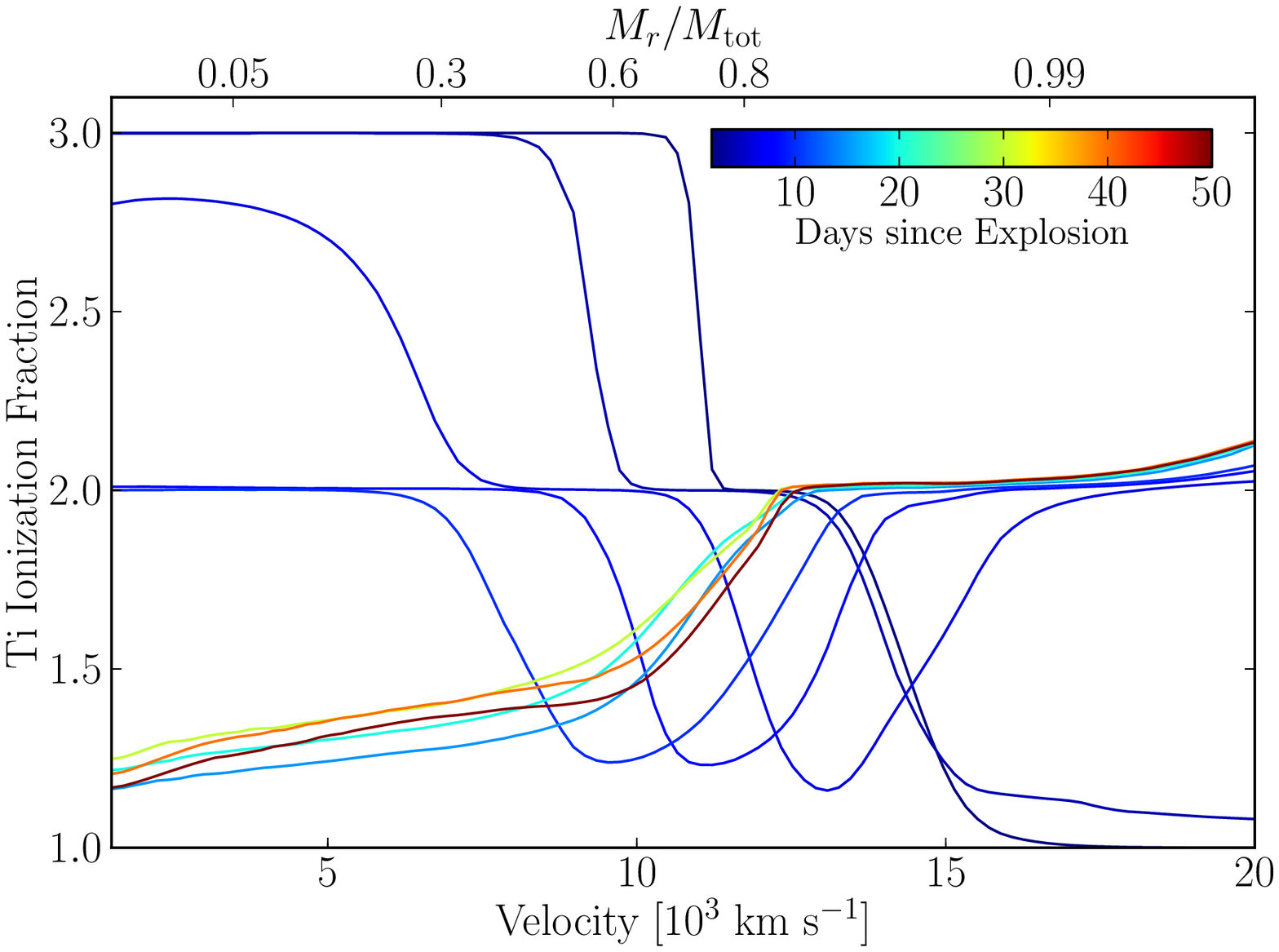,width=8.5cm}
\caption{Same as Fig.~\ref{fig_prop_evol}, but now showing
the helium, calcium, and titanium ionization states at the photosphere
(a value of zero corresponds to a neutral state, of value of $i$ to an ith-ionized atom).
Note that the ordinate range varies amongst panels.
The  times plotted are 2, 4, 6, 8, 10, 15, 20, 30, 40, and 50\,d after explosion.
\label{fig_ionfrac_evol}
}
\end{figure*}

We perform slight adjustments to the original ejecta produced by the hydrodynamical simulation \citep{waldman_etal_11}.
We smooth the density and elemental distribution
to reduce sharp gradients,\footnote{The smoothing is obtained by convolving the corresponding distributions
with a gaussian whose standard deviation is 500\,\kms.} which are largely an artifact of the 1-D treatment of the explosion.
To ensure the outer boundary is optically thin at early times,
we also replace and extend the density structure beyond 16000\,\kms\ by a power law in velocity with an exponent of $-12$
(the original density profile is poorly resolved and noisy at large velocities).
The radiative-transfer simulations for model COp45HEp2 are started at 10$^5$ seconds after explosion
and employ 100 grid points.
The ejecta is in homologous
expansion and covers velocities from 1100 to 29000\,\kms\ (the outer radius is thus initially at 2.9$\times$\,10$^{14}$\,cm).
The outer ejecta temperature (mass density) is $\sim$\,2000\,K ($\sim$10$^{-15}$\,g\,cm$^{-3}$)
so that the outer grid is transparent to radiation. Further details on the COp45HEp2 model properties at 1.16\,d can
be found in \citet{waldman_etal_11}.

In \cmfgen, the radiative transfer solver used is generally the same whatever the SN ejecta under study.
What differs are the initial conditions  (mass, density, explosion energy, composition etc.)
and the model atoms employed.  For this work, we use the model
atom set named A3 in \citet{D14}, with several adjustments. We add He\one\ and He\two, as well as neutral states
of Mg, Si, S, and Ca. We also employ a large Ti\two\ model atom with 1000 levels (61 super levels).
We treat nine two-step decay chains, whose leading isotope are \iso{56}Ni, \iso{57}Ni, \iso{48}Cr, \iso{49}Cr,
\iso{51}Mn, \iso{55}Co, \iso{37}K, \iso{52}Fe, and \iso{44}Ti. The characteristics of these chains are summarized
in \citet{D14}. Non-local energy deposition is allowed for once the model reaches an age of 3\,d --- prior to
that time we assume all the radioactive energy is deposited at the site of decay.
However, at all times, we assume that positrons are destroyed locally, and hence their kinetic energy 
is deposited locally, i.e., at the site of emission.
This is done for convenience since we lack late-time observations to constrain adequately the positron trapping 
efficiency. This efficiency should be high over the 1-2 months time scale considered here, although
positrons injected through \iso{44}Sc decay may increasingly escape, in the absence of a magnetic field, 
as the ejecta ages \citep{milne_etal_99,perets_14}.

There are several deficiencies in our models related to the atomic data. In particular,
we do not have accurate photoionization cross-sections for Ti, and this could potentially impact the influence
of Ti{\, \sc ii} on the spectra through an incorrect ionization balance. Further, in most of our modeling
we did not allow for the influence of Vanadium on the spectrum -- we simply included one level of V\one\
in order to track the changes in composition associated with the  \iso{48}Cr $\rightarrow$ \iso{48}V $\rightarrow$ \iso{48}Ti
decay chain. Limited testing with newly developed Ti and V atoms show only limited changes of the predicted spectra.

\begin{figure*}
\epsfig{file=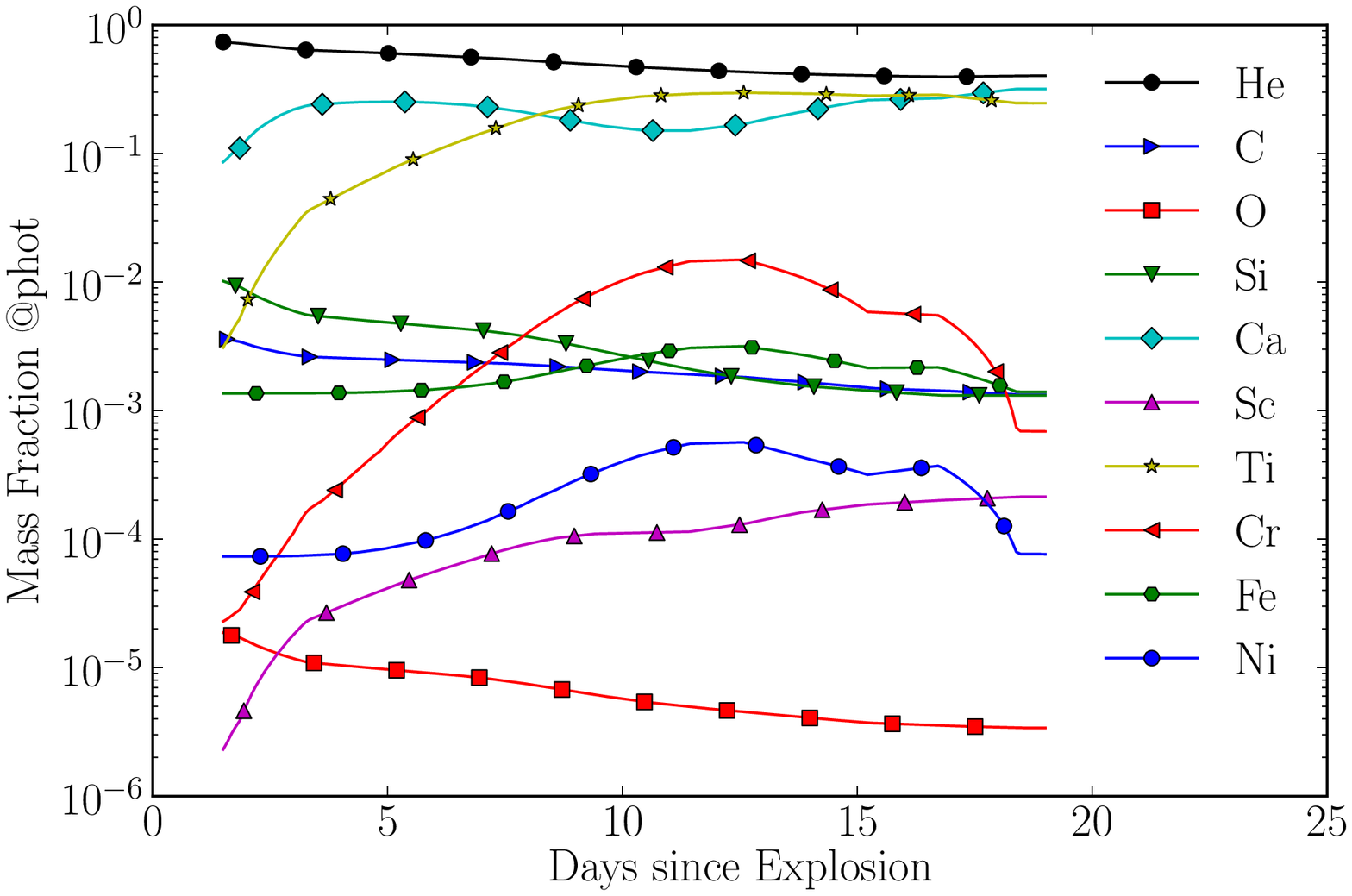,width=8.5cm}
\epsfig{file=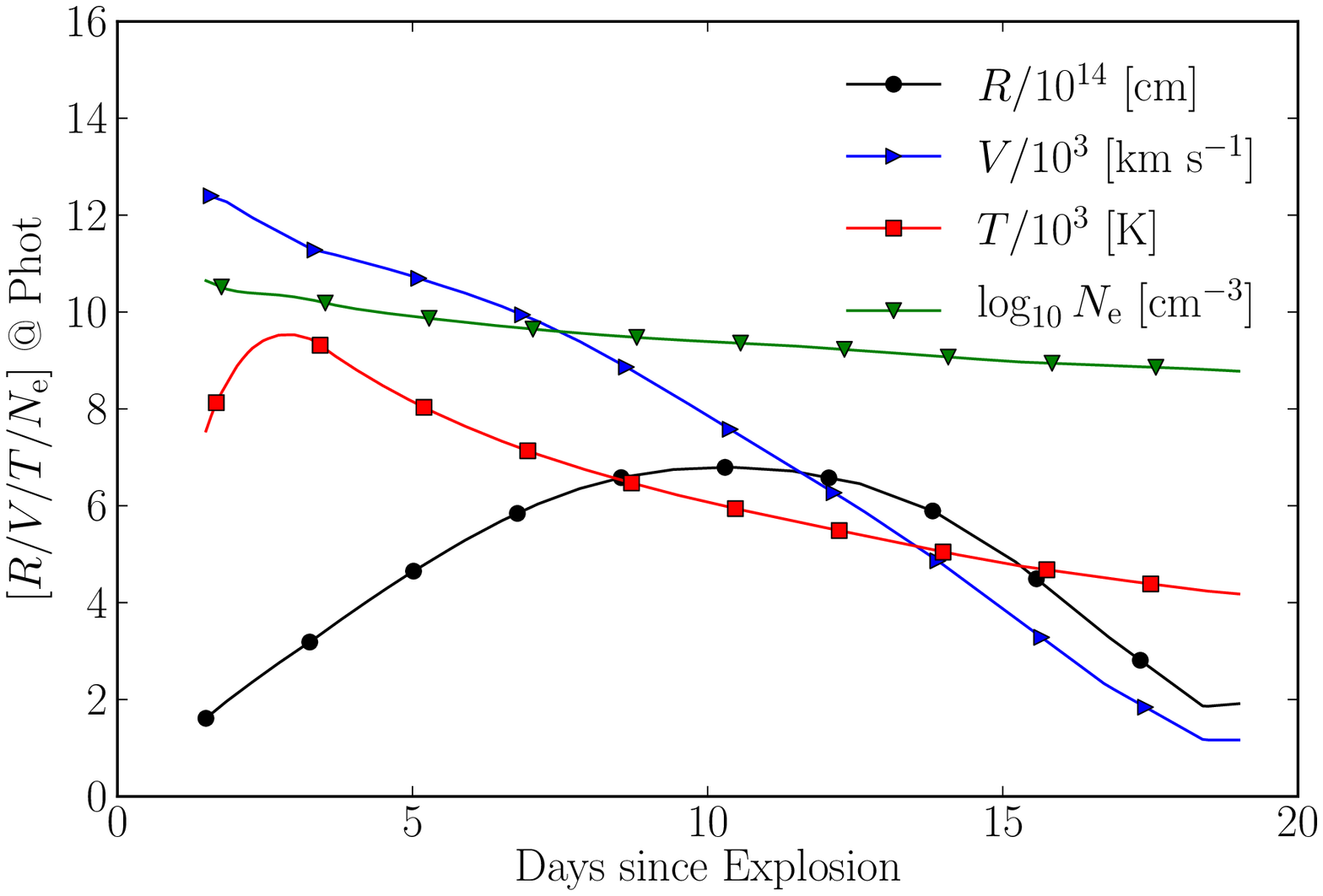,width=8.5cm}
\caption{{\it Left:} Evolution at the photosphere of the mass fraction for He, C, O, Si, Ca, Sc, Ti, Cr, Fe, and Ni.
We define the photosphere as the ejecta location where the inward-integrated Rosseland-mean
optical depth is 2/3. The photosphere, defined that way, reaches the base of the ejecta at 18.4\,d
after explosion. 
{\it Right:} Same as left, but now showing the evolution of the radius, velocity, temperature, and
electron density (with appropriate scalings) at the photosphere as a function of time.
If we use electron scattering instead of the Rosseland-mean opacity for the definition of the photosphere,
these curves are shifted by $\sim$\,10\,\%.
\label{fig_comp_phot}
}
\end{figure*}

\section{Power source for radiation}
\label{sect_power}

Immediately after explosion, the most abundant unstable isotopes in model COp45HEp2
are \iso{44}Ti, \iso{48}Cr, \iso{52}Fe, and \iso{56}Ni (shown as dashed lines
in Fig.~\ref{fig_comp_init}), with initial  masses of 0.033, 0.0056, 0.0009, and 0.0001\,\msun, respectively.
The isotope \iso{44}Ti, which decays to \iso{44}Ca via \iso{44}Sc, is the primary constituent of Ti initially.
Despite the large abundance, its very large half life of 21915 days makes it a secondary power source
during the first 1-2 months after explosion.
Conversely, the very short half lives (0.34479 and 0.01472 days) of  \iso{52}Fe and its daughter isotope \iso{52}Mn
make them weak power sources beyond a day.
In our ejecta model, the main power source is the chain \iso{48}Cr $\rightarrow$  \iso{48}V $\rightarrow$  \iso{48}Ti.
\iso{48}Cr has  a half-life of 0.89833\,d and releases 0.430\,MeV per decay while
\iso{48}V has a half-life of 15.9735\,d and releases  3.0553\,MeV (of which 0.1449\,MeV is positron energy) per decay.
The energy per decay is comparable to the \iso{56}Ni chain (total of 5.4671\,MeV),
 but with a shorter characteristic time. However, the low mass of  \iso{48}Cr in this ejecta can only generate
 a small power compared to SNe Ia, which have approximately 100 times more \iso{56}Ni.

\section{Results: Ejecta properties}
\label{sect_ejecta}

 Of particular interest are the temperature, the ionization state, and the electron
density because they control the ejecta optical depth and the ions present at the photosphere.
Figures~\ref{fig_prop_evol} and \ref{fig_ionfrac_evol} illustrate these quantities.

The ejecta optical depth, temperature and ionization are inter-related. For as long as the temperature
stays above $\sim$\,10000\,K (Fig.~\ref{fig_prop_evol}), helium, which is the dominant species at all ejecta locations, remains
partially (or fully) ionized (Fig.~\ref{fig_ionfrac_evol}). Early on, the separation between ionized and partially
neutral regions is the photosphere --- it coincides with the jump in electron density and in helium ionization
(Fig.~\ref{fig_ionfrac_evol}).
When the ejecta temperature drops everywhere below $\sim$\,10000\,K,
helium recombines to its neutral state, the mass absorption coefficient (here, we use the Rosseland-mean
opacity, which becomes merely indicative when the conditions turn optically thin)
suddenly decreases, and the  ejecta optical depth drops below unity.
The opacity is also function of the density and the temperature (which affects collisional processes, level populations etc.)
but the electron density (and the process of electron scattering) is key for the gas opacity.
At 2\,d after explosion, the outer 30\,\% of the ejecta is neutral and optically thin to radiation.
Assuming a constant opacity with depth and time is therefore inaccurate. Although the temperature
continuously decreases at velocities less than 11000\,\kms, non-thermal processes maintain a
persistent He$^+$ population with an ionization fraction of $\sim$\,5\% or larger.

When the conditions are ionized, we find that non-thermal processes play little or no role and the entire decay
energy is deposited as heat \citep{xu_mccray_91,dessart_etal_12,li_etal_12}. At and above the photosphere
(and throughout the ejecta when it turns thin), the fraction of the decay energy channeled into
non-thermal ionization and excitation at the photosphere is $\sim$\,20\,\% and 10\,\%, respectively.
As discussed in Section~\ref{sect_spec}, this causes the production of He\one\ lines at light curve peak.

The evolution of the photospheric properties is shown in Fig.~\ref{fig_comp_phot}.
The photospheric mass fraction of He and of IGEs evolves little with time, but increases
by 1-2 orders of magnitude for IMEs, especially for Sc, Ti, and Cr. This evolution is primarily
caused by the chemical stratification rather than radioactive decay.
The photosphere reaches a maximum radius of 6$\times$10$^{14}$\,cm at 10\,d. Prior to
that, the photosphere moves at a velocity in excess of 7000\,\kms; this is
in the range of photospheric velocities seen in Ca-rich transients at early times \citep{perets_etal_10,kasliwal_etal_12}.
Photospheric temperatures are in the range 4000--8000\,K and will cause the presence of lines from
neutral and once-ionized species in spectra. The most abundant species at the photosphere
(in fact, at all depths and thus at all times) are He, Ca, and Ti.

\begin{figure*}
\epsfig{file=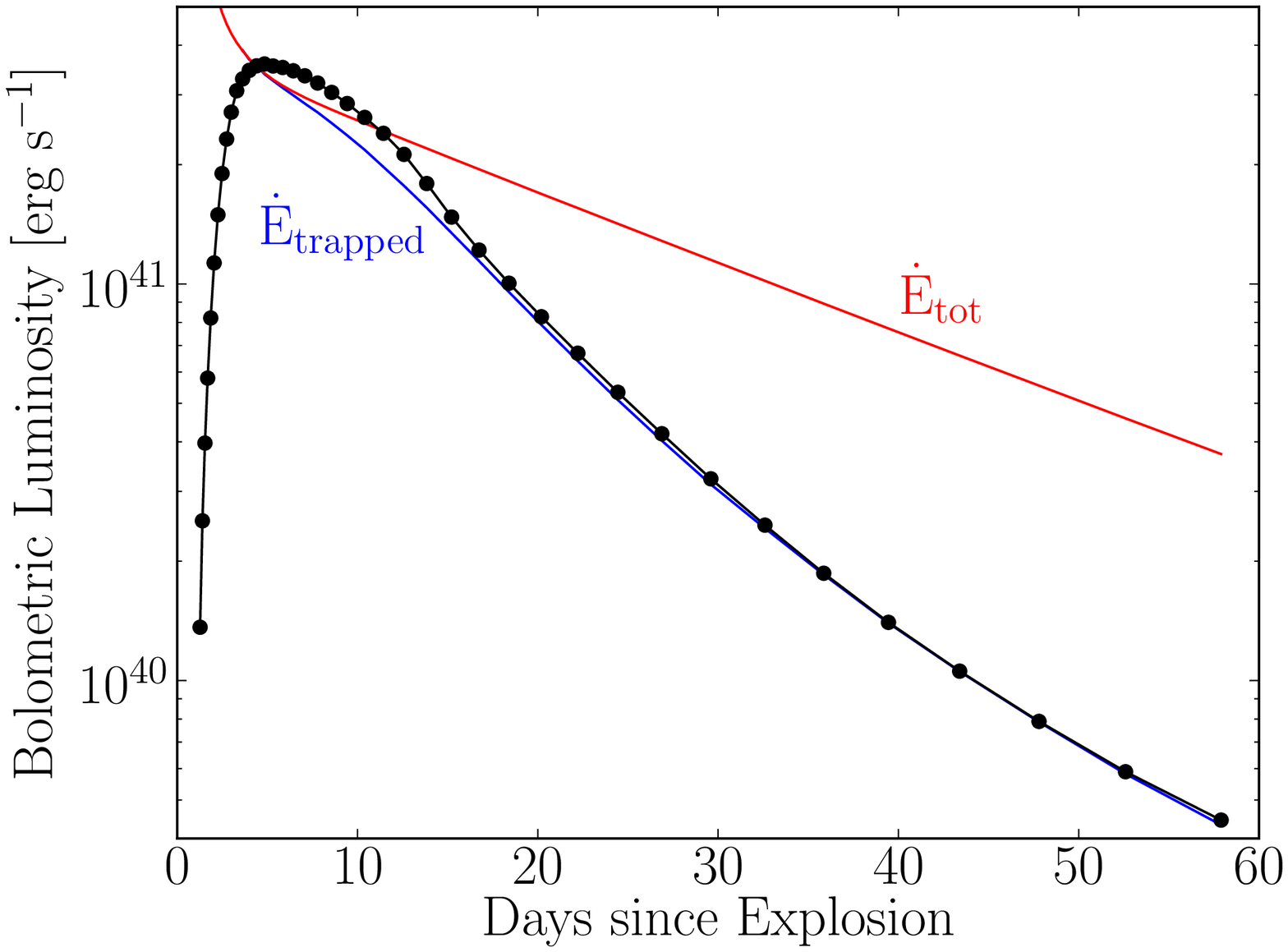,width=8.5cm}
\epsfig{file=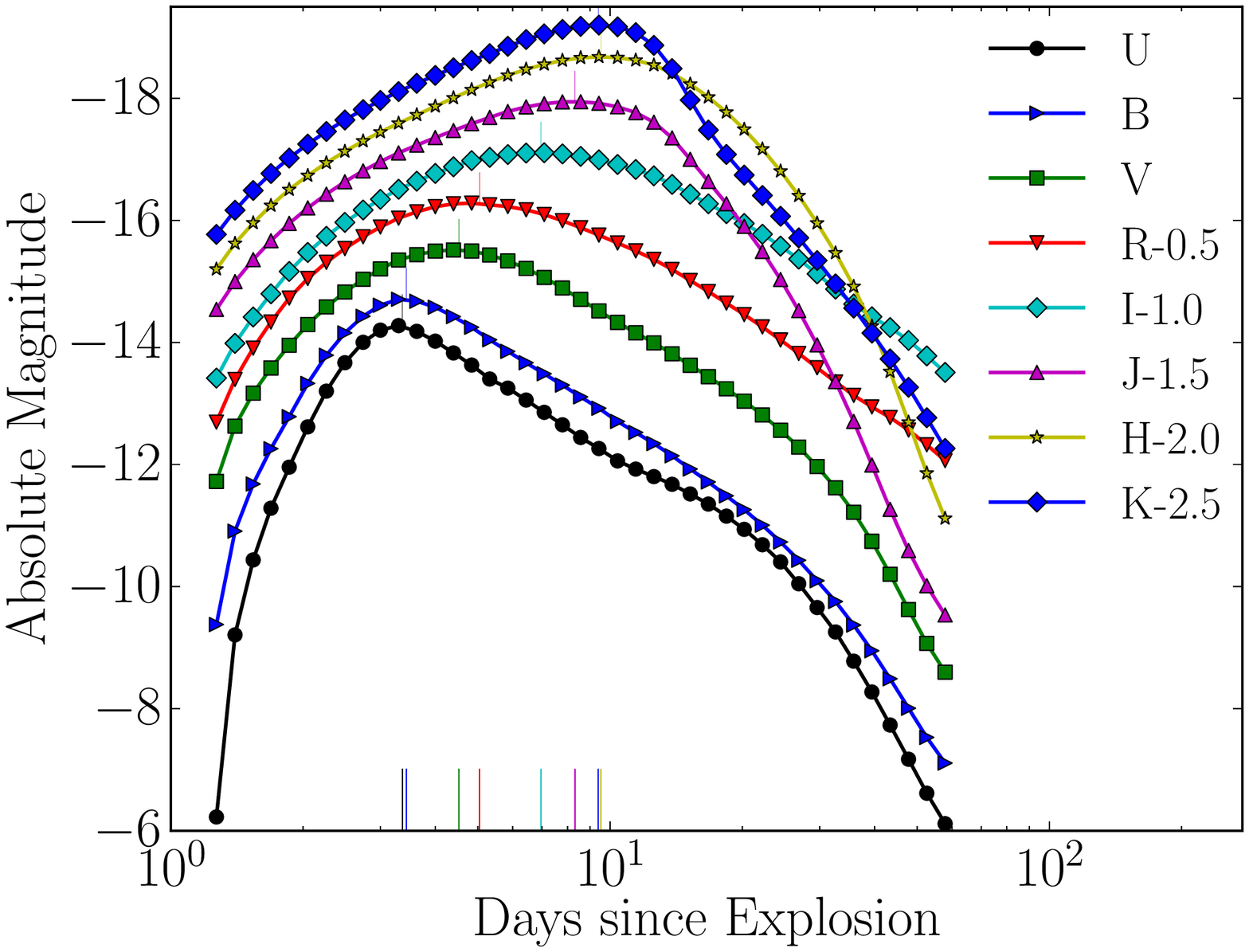,width=8.5cm}
\caption{
{\it Left:} Bolometric luminosity of model COp45HEp2. We also add the total decay energy power (red) and the fraction
that is trapped within the ejecta (blue). The time of bolometric maximum is 4.84\,d after explosion.
{\it Right:} Same as left, but now for the multi-band light curves. Tick marks at bottom indicate the rise time in each filter.
In contrast, standard SNe Ia have a bolometric maximum of $\sim$\,10$^{43}$\,erg\,s$^{-1}$ at $\sim$\,18\,d after
explosion, and exhibit a secondary maximum in near-IR bands.
\label{fig_mod_lc}
}
\end{figure*}

\begin{figure*}
\epsfig{file=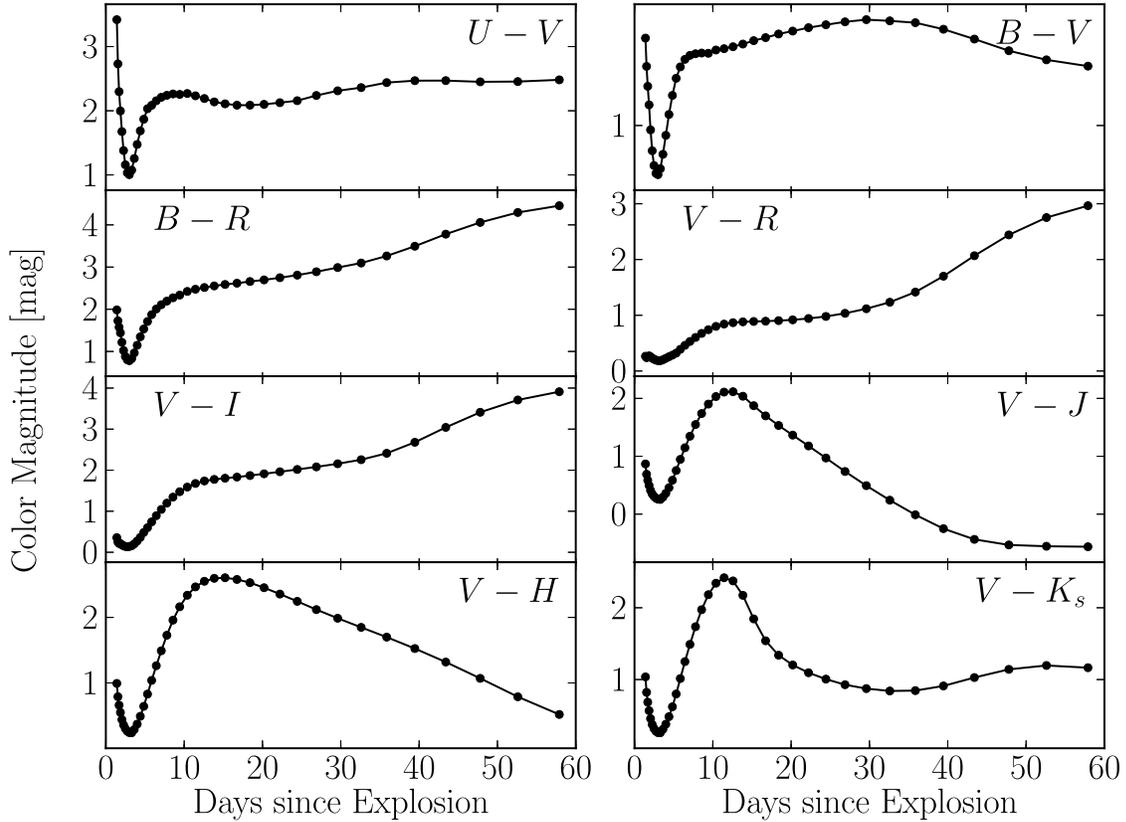,width=15cm}
\caption{Illustration of the color evolution of model COp45HEp2 for different filter combinations.
\label{fig_colours}
}
\end{figure*}

\section{Results: Bolometric luminosity and multi-band light curves}
\label{sect_phot}

   The bolometric light curve of model COp45HEp2 shows a very rapid evolution, with a peak
of 3.59$\times$10$^{41}$\,erg\,s$^{-1}$ at only 5.17\,d after explosion (left panel of Fig.~\ref{fig_mod_lc}).
For comparison, standard-energy
delayed-detonation models of Chandrasekhar-mass WDs give a bolometric maximum of
$\sim$10$^{43}$\,erg\,s$^{-1}$ and a rise time of $\sim$18\,d \citep{blondin_etal_13}.
These differences stem from the greater decay energy released in SNe Ia (the mass of \iso{56}Ni is 100 times greater
in standard SNe Ia than the mass of \iso{48}Cr in model COp45HEp2)
and the larger ejecta mass in SNe Ia (1.4 compared to 0.2\,\msun\ in model COp45HEp2). The
energy to mass ratio is however comparable between a typical SN Ia and the He-shell detonation
model COp45HEp2.

At maximum, the luminosity is comparable to the decay power (Fig.~\ref{fig_mod_lc}),
as expected \citep{arnett_82} and obtained for SNe Ia \citep{hoeflich_khokhlov_96, blondin_etal_13}.
Past maximum, it exceeds the decay power until it turns optically thin at about two weeks
after explosion. This excess stems from stored energy within the ejecta and remains visible
as long as the energy can be trapped.  The model luminosity around light curve maximum is powered
at the $\gtrsim$95\,\% level by the decay of \iso{48}Cr and its daughter element \iso{48}V.
The escape of $\gamma$ rays starts around maximum  (blue versus red curve) and causes
a large leakage of energy --- by 60\,d after explosion only $\sim$\,10\% of the total decay energy is trapped.

   These properties are consistent with the heuristic estimates of \citet{bildsten_etal_07}. We find a
   bolometric maximum 30\,\% higher and a rise time that is 30\% shorter than obtained by
   \citet{waldman_etal_11} for  that same model. The reason
   for this is probably that they adopt a gray opacity, approximate its dependency on composition,
   and ignore the influence of ionization. Still, the key feature of a very short rise time to a modest
   maximum bolometric luminosity remains. In general, adopting a fixed mass absorption
   coefficient or applying simple scalings based on SN Ia ejecta \citep{perets_etal_10}
   will probably lead to an overestimate of the opacity and an underestimate of the ejecta mass
   (see Section~\ref{sect_comp}).

   Multi-band light curves show large variations between filters (right panel of  Fig.~\ref{fig_mod_lc}).
Rise times to maximum increase monotonically with the mean wavelength of each filter.
The times of maximum magnitude are
3.37 ($U$), 3.44 ($B$), 4.53 ($V$), 5.04 ($R$), 7.07 ($I$), 8.38 ($J$), 9.62 ($H$), and 9.37\,d ($K$).
The fading rate past maximum varies with time and differs significantly between filters.
For example, the magnitude fading between maximum and 15 days later covers the large range
from 1.3 to $\gtrsim$\,3\,mag (Table~\ref{tab_lc}).
Interestingly, the near-IR bands show no secondary maximum, in contrast to standard SN Ia
explosions. 
There are probably several reasons for this, including the difference in temperature evolution
(our model is cool and has red colors even at peak), in ionization (Ca and Ti remain once 
ionized after bolometric maximum), and in composition (the Co intimately related to the secondary 
maximum of SNe Ia is strongly under-abundant in our COp45HEp2 model). We do not dwell further upon
this issue since there is at present no near-IR observation (neither photometric nor spectroscopic) of 
.Ia SNe and Ca-rich transients.

       During the photospheric phase (i.e. prior to $\sim$\,20\,d), the color evolution of model COp45HEp2
in the optical and near-IR is towards bluer colors up to the peak and then redder (Fig.~\ref{fig_colours}).
When the ejecta turns thin (which occurs around 18\,d after explosion, depending on what opacity
is used for the inference), the color evolution differs between filters. For example, $V-H$ decreases again (color shift to blue),
while $U-V$ is roughly constant and $B-R$ increases (color shift to red). This complex behavior is
related to the presence of strong emission lines (primarily forbidden) whose wavelength distribution
coincidently overlap, or not, with specific filters. We discuss these spectral characteristics in the
next section.

\begin{table*}
\begin{center}
    \caption{Summary of photometric properties for model COp45HEp2. For each entry, we give the
 rise time to maximum, the value at maximum, and the magnitude change between peak and 15\,d later
 for the corresponding filter band. Numbers in parentheses are powers of ten.
\label{tab_lc}}
\begin{tabular}{c@{\hspace{3mm}}
c@{\hspace{3mm}}c@{\hspace{3mm}}c@{\hspace{3mm}}
c@{\hspace{3mm}}c@{\hspace{3mm}}c@{\hspace{3mm}}
c@{\hspace{3mm}}c@{\hspace{3mm}}c@{\hspace{3mm}}
c@{\hspace{3mm}}c@{\hspace{3mm}}c@{\hspace{3mm}}
}
\hline
 & \multicolumn{3}{c}{$L_{bol}$}   & \multicolumn{3}{c}{$L_{UBVRI}$}   & \multicolumn{3}{c}{$U$}   & \multicolumn{3}{c}{$B$} \\
\hline
 & $t_{\rm rise}$ & Max. & $\Delta M_{15}$ & $t_{\rm rise}$ & Max. & $\Delta M_{15}$ & $t_{\rm rise}$ & Max. & $\Delta M_{15}$ & $t_{\rm rise}$ & Max. & $\Delta M_{15}$ \\
 & [d] & erg\,s$^{-1}$ & mag  & [d]  & erg\,s$^{-1}$ & mag    & [d]  &mag & mag  & [d]  & mag & mag \\
\hline
 & 5.171(0)  &   3.590(41)  &  1.591(0)  &   4.330(0)  &   2.640(41)  &  1.714(0)  &   3.367(0)  &  -1.428(1)  &   3.124(0)  &   3.438(0)  &  -1.472(1)  &   3.236(0) \\
\hline
& \multicolumn{3}{c}{$V$}   & \multicolumn{3}{c}{$R$}   & \multicolumn{3}{c}{$I$}  & \multicolumn{3}{c}{}   \\
\hline
 & $t_{\rm rise}$ & Max. & $\Delta M_{15}$ & $t_{\rm rise}$ & Max. & $\Delta M_{15}$ & $t_{\rm rise}$ & Max. & $\Delta M_{15}$ &  & &  \\
                & [d]  &mag & mag  & [d]  &mag & mag   & [d]  & mag & mag  \\
\hline
 & 4.526(0)  &  -1.552(1)  &   2.401(0)  &   5.044(0)  &  -1.578(1)  &   1.807(0)  &   7.065(0)  &  -1.610(1)  &   1.318(0) & & & \\
\hline
 & \multicolumn{3}{c}{$J$}   & \multicolumn{3}{c}{$H$}   & \multicolumn{3}{c}{$K$}  & \multicolumn{3}{c}{}   \\
\hline
 & $t_{\rm rise}$ & Max. & $\Delta M_{15}$ & $t_{\rm rise}$ & Max. & $\Delta M_{15}$ & $t_{\rm rise}$ & Max. & $\Delta M_{15}$ &  & &  \\
                & [d]  &mag & mag  & [d]  &mag & mag   & [d]  & mag & mag  \\
  &  8.378(0)  &  -1.644(1)  &   2.694(0)  &   9.622(0)  &  -1.668(1)  &   1.904(0)  &   9.369(0)  &  -1.669(1)  &   3.116(0)  & & & \\
\hline
\end{tabular}
\end{center}
\end{table*}

\section{Results: Spectra}
\label{sect_spec}

\begin{figure*}
\vspace{-0.5cm}
\epsfig{file=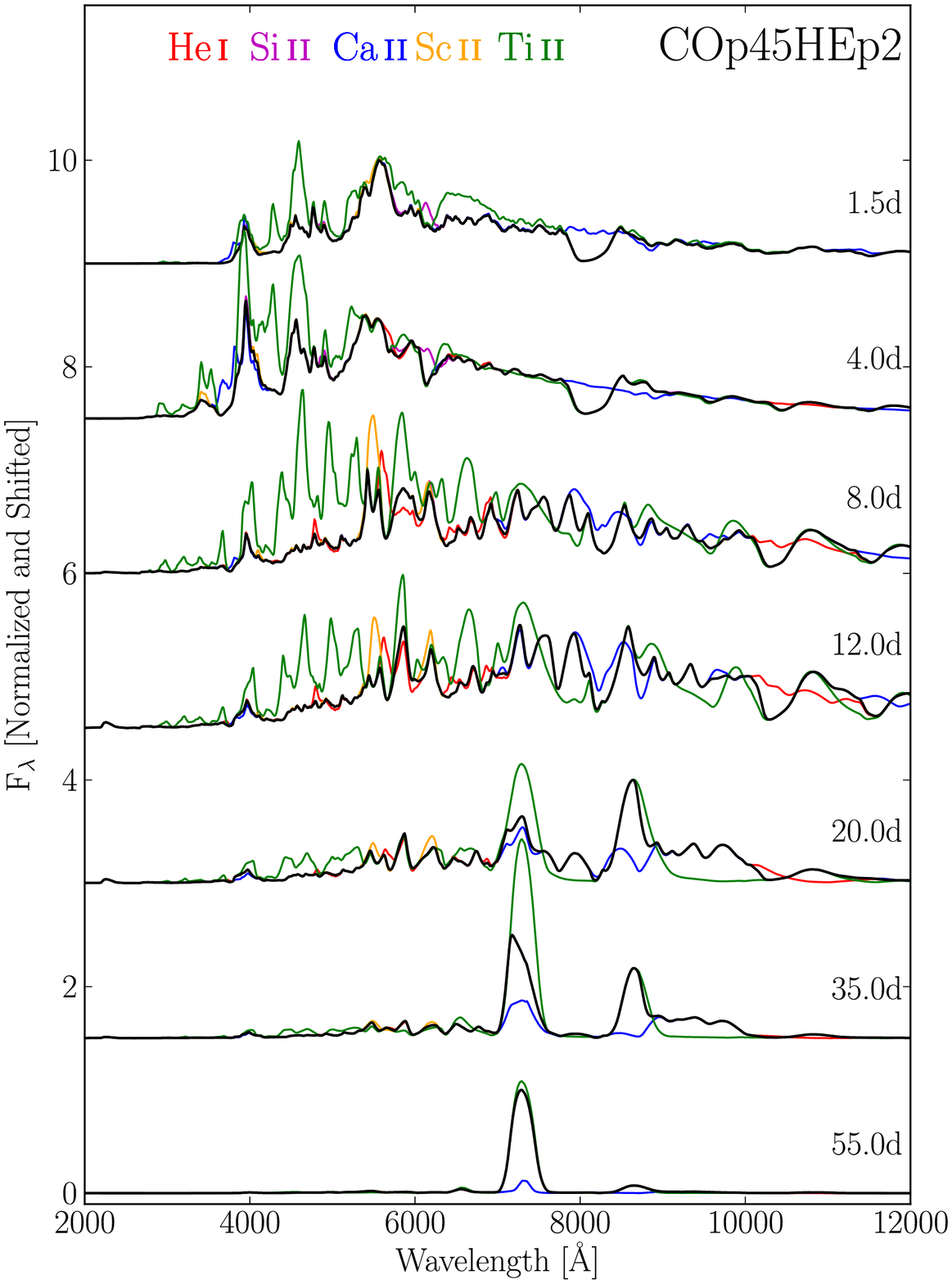,width=8.75cm}
\epsfig{file=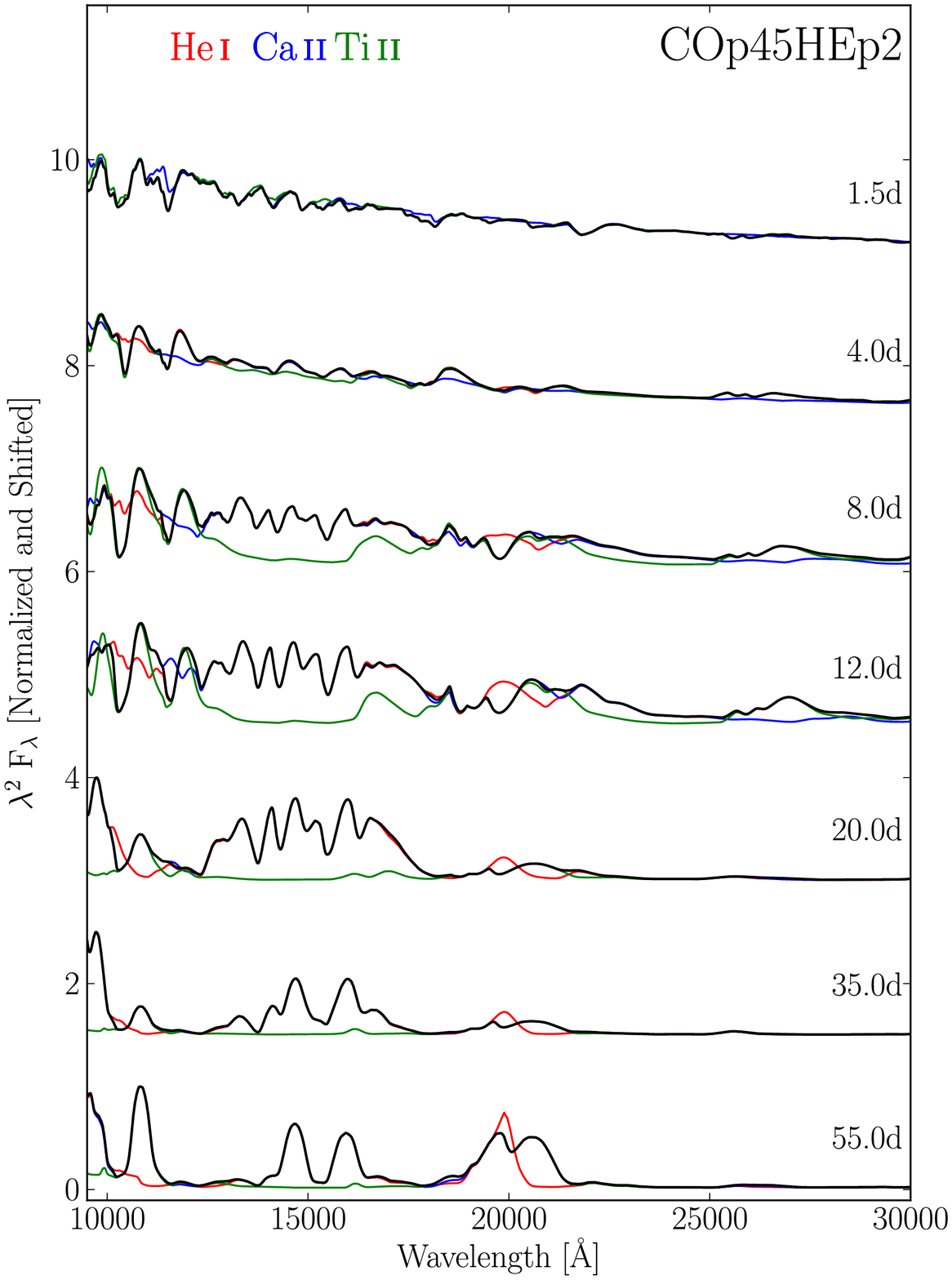,width=8.75cm}
\vspace{-0.8cm}
\caption{
{\it Left:} Montage of synthetic spectra for model COp45HEp2 over the optical range, from 1.5 until 55\,d.
To facilitate the interpretation of the spectral features, we add
synthetic spectra (color) where the bound-bound transitions of a selected ion (see label at top)
is excluded from the spectrum calculation.
A vertical shift of 1.5 is applied to each epoch (the flux at $\sim$\,2000\,\AA\ is close to zero at all times).
{\it Right:} Same as leff, but now showing the quantity $\lambda^2 F_{\lambda}$ in the near-IR range.
\label{fig_spec_model}
}
\end{figure*}

   A striking property of model COp45HEp2 is the unambiguous presence of He\one\ lines in the
   optical and near-IR ranges around maximum brightness (i.e., from 5 to 15\,d after explosion; Fig.~\ref{fig_spec_model}).
   At maximum brightness, the He mass fraction at the photosphere is 0.6 (Fig.~\ref{fig_comp_phot}).
   As in standard SNe Ib, He\one\ lines arise from efficient non-thermal excitation and ionization \citep{lucy_91}.
   In the optical, we see He\one\ lines at 5875.66 (3d\,$^3$D --\,2p\,$^3$P$^{\rm o}$),
   6678.15 (3d$^1$D--\,2p$^1$P$^{\rm o}$), and 7065.25\,\AA\ (3s$^3$S--\,2p$^3$P$^{\rm o}$).
   Si\two\,6347--\,6371\,\AA\ (4p\,$^2$P$^{\rm o}$--\,4s\,$^2$S doublet)
   causes a strong feature up until bolometric maximum (i.e., $\sim$\,4\,d)
   and therefore produces a unique classification to this event as both a SN Ib and a SN Ia.
   There is also a broad absorption around 5000-5600\,\AA\ caused by S\two\ (numerous
   transitions associated with the term 4p--4s). The same features
   affect the spectra of SNe Ia (there, the Si and S abundances are higher but the lines are typically
   saturated).   No line of O\one\ is seen anywhere in the model spectrum --- the O mass fraction is
   only $\sim 2\times 10^{-6}$.  A test calculation with the oxygen abundance increased by a factor of 100 had no effect.

   Calcium shows the usual strong optical lines with Ca\two\,H\&K (4s\,$^2$S--\,4p\,$^2$P$^{\rm o}$ terms),
   the triplet lines around 8500\,\AA\ (3d\,$^2$D--\,4p\,$^2$P$^{\rm o}$ terms),
   and the forbidden transitions at 7307\,\AA\ (4s\,$^2$S--\,3d\,$^2$D terms).
   These forbidden lines appear as early as 20\,d after explosion in our simulations,
   just as the Rosseland-mean optical depth  drops below unity.
   Scandium contributes some moderate blanketing (below $\sim$\,10000\kms, its mass fraction
   is up by four orders of magnitude above
   the solar metallicity value). Two strong and broad features are unambiguously
   seen, at 5600\,\AA\ (terms 3p$^6$\,3d\,4p\,$^3$P$^{\rm o}$--\,3p$^6$\,3d$^2$\,$^3$P)
   and at 6260\,\AA\ (terms 3p$^6$\,3d\,4p\,$^3$D$^{\rm o}$--\,3p$^6$\,3d$^2$\,$^3$P).

   With an ejecta mass fraction everywhere above $\sim$\,0.1, titanium has the strongest impact on the spectrum.
   In the optical it causes significant blanketing  in the blue and emission in the red, and is seen over the first month.
   It affects other strong spectral features, causing significant absorption that overlaps with He\one\,5875\,\AA\
   and the forbidden doublet of Ca\two. It can also lead to confusion with O\one\ --  the feature at $\sim$\,7700\,\AA\
   is due to Ti\two\ rather than O\one. The forest of Ti\two\ lines in the optical corresponds to thousands of transitions
   that are not possible to list. For example, strong transitions occur between upper levels
   3d$^2$($^3$F)\,4p\,z$^4$D$^{\rm o}$  (z$^2$P$^{\rm o}$ etc.) and lower levels like 3d$^2$($^3$P)\,4s\,b$^4$P.
      Although Ti is not considered an IGE, its complex atomic structure can cause dramatic line blanketing.
   Ti\two\ line blanketing is so strong that it overwhelms the negligible blanketing associated with the much less
   abundant Fe\two. Combined with the modest temperature of the photosphere, it causes the red color of the
   model at all times, even around bolometric maximum.

   The near-IR range contains fewer spectral features (see right panel of Fig.~\ref{fig_spec_model}).
   We obtain strong He\one\,10830\,\AA\
   and a weaker He\one\,20581\,\AA. Ca\two\,(5s$^2$S--5p$^{2}$P$^{\rm o}$)
   gives a strong feature at 11839\,\AA.
   All other lines are due to Ti\two, in particular forbidden transitions
   associated with upper levels 3d$^2$($^3$F)\,4p\,z$^4$F$^{\rm o}$ (and z$^2$D$^{\rm o}$)
   and lower levels 3d\,4s$^2$\,c$^2$D (3d$^3$\,b$^2$F) with wavelengths
  10113.87, 13320.45, 14151.54, 14631.55, 14711.64, 15231.10, and 15873.64\,\AA.
  Interestingly, it is the strengthening and weakening of these lines that control
  the near-IR photometry. In particular, the extended brightness in the $H$-band (which
  encompasses the range 1.5--1.8\,$\mu$m)
  up to 20\,d after explosion is caused by [Ti\two]. In such ejecta, the
  near-IR range is the ideal place to search for signatures of Ti, determine its abundance,
  and quantify the extent to which it contributes to the \iso{44}Ca enrichment of the ISM
  \citep{mulchaey_etal_14}.

 \begin{figure*}
\epsfig{file=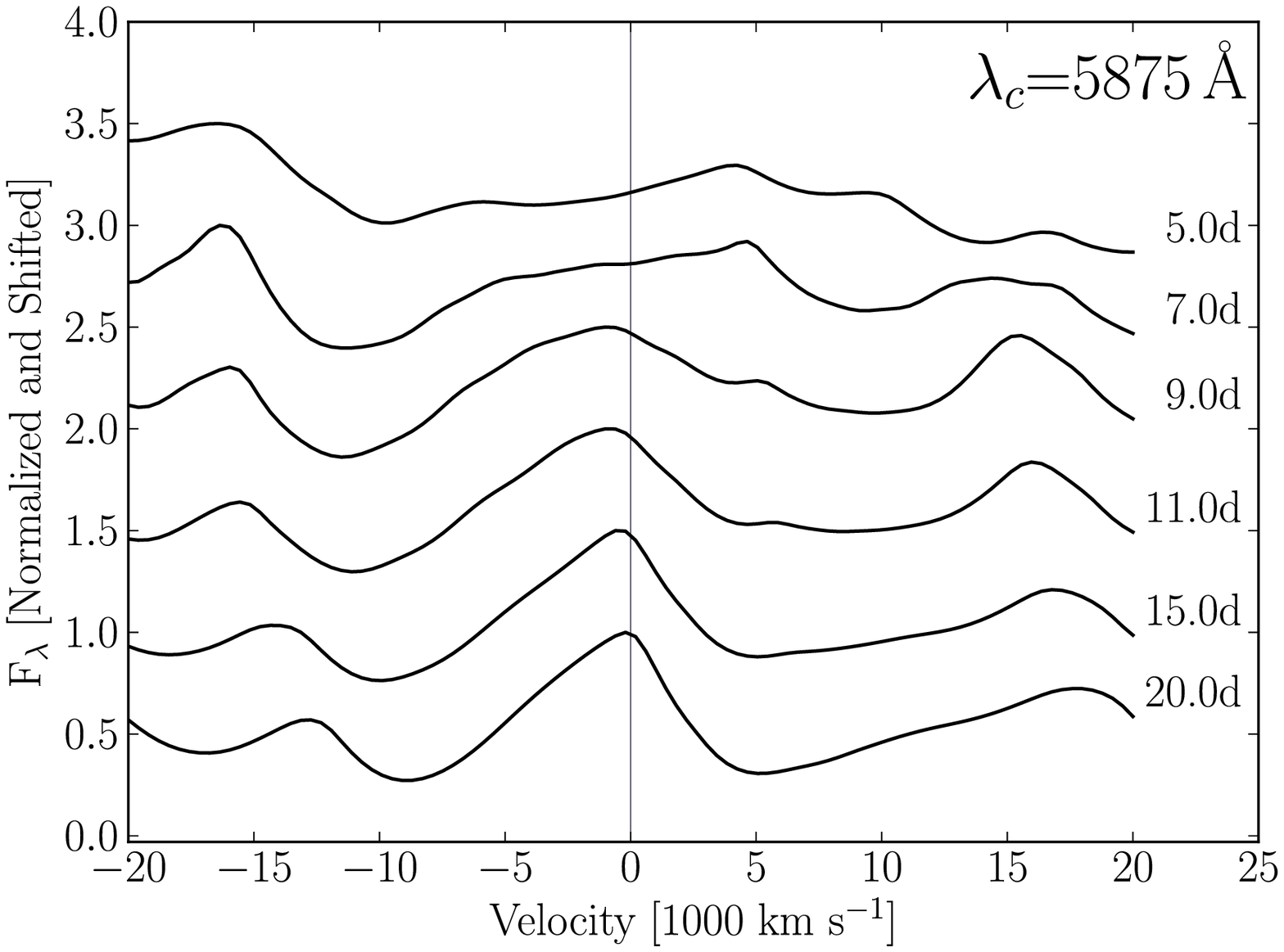,width=8.75cm}
\epsfig{file=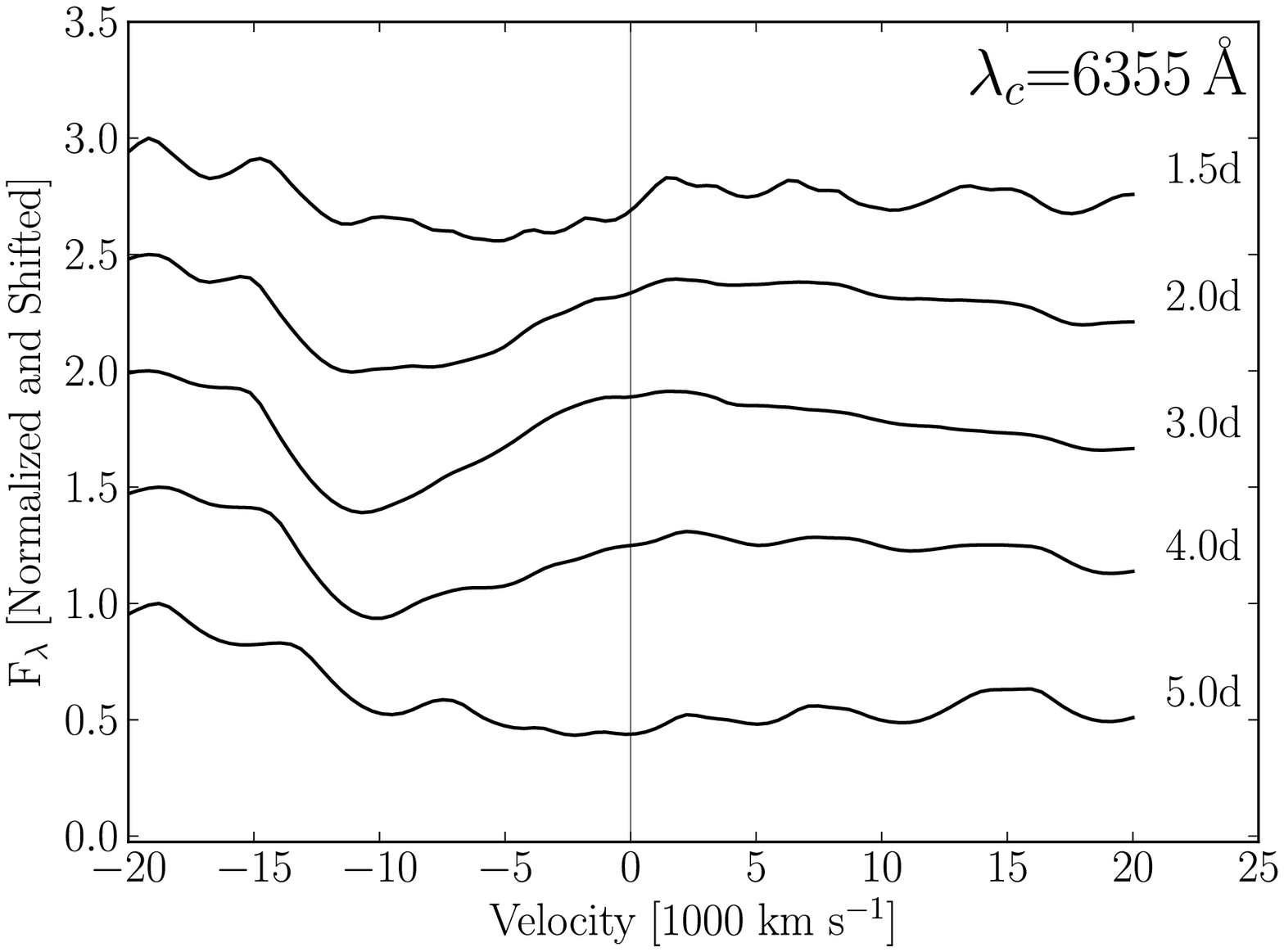,width=8.75cm}
\epsfig{file=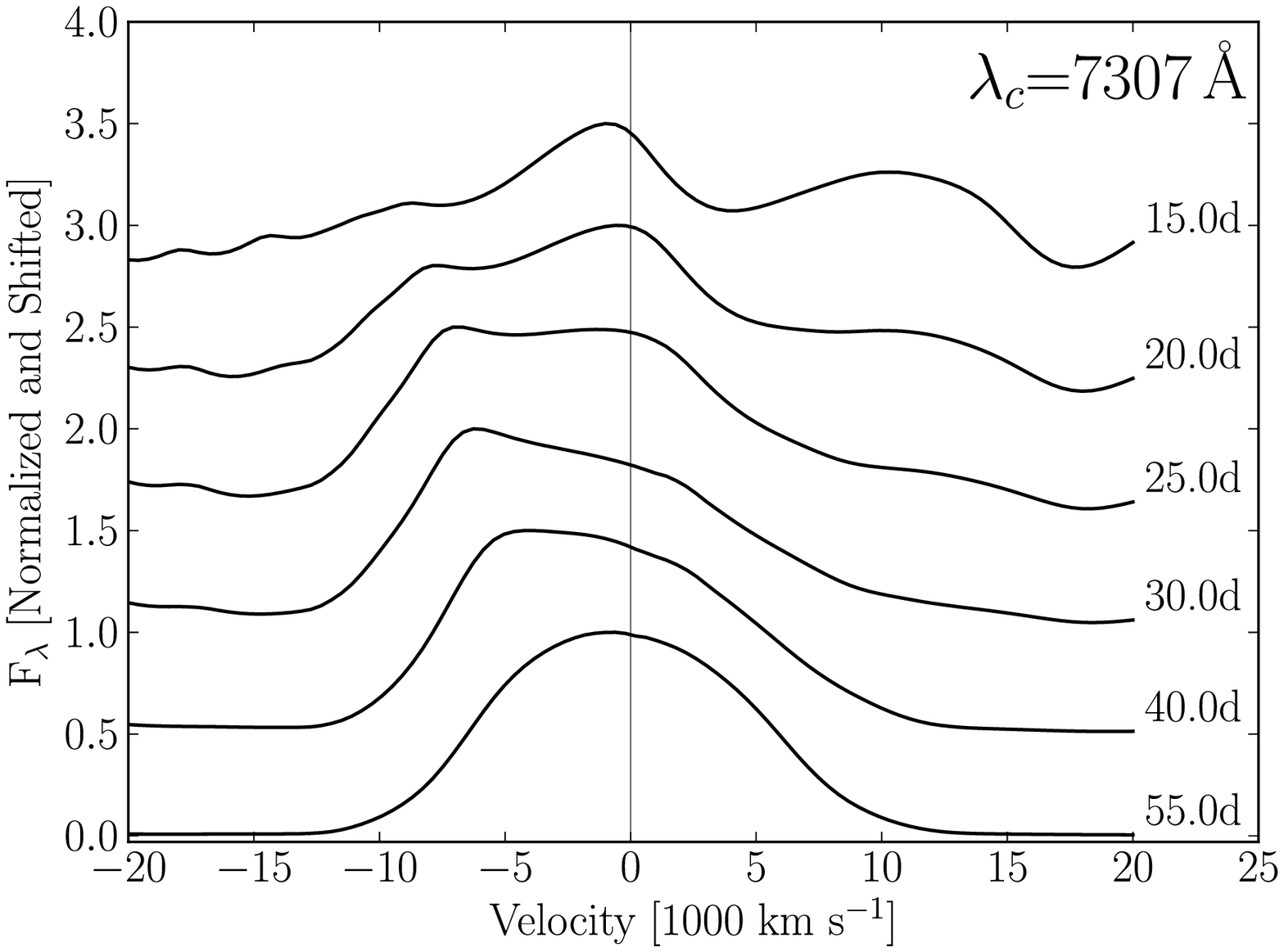,width=8.75cm}
\epsfig{file=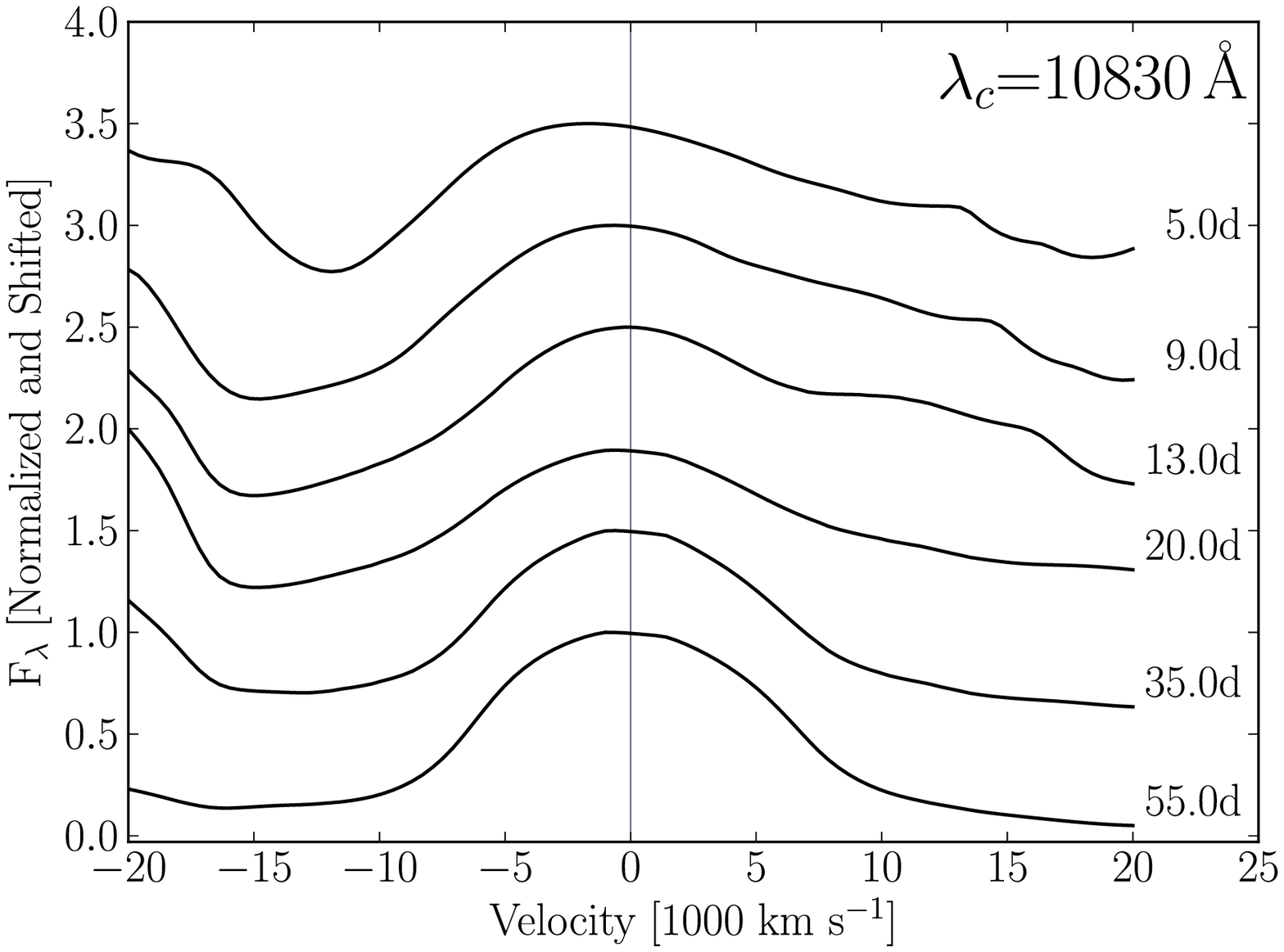,width=8.75cm}
\caption{
Montage of synthetic spectra for model COp45HEp2, in velocity space, and focusing on
specific spectral regions. He\one\,5875\,\AA\ is present over the epochs 5-15\,d, and becomes
more and more affected by Ti\two\ line absorption (top left).
Si\two\,6355\,\AA\ is only visible for a few days after explosion (top right).
Ca\two\,7307\,\AA\ develops into a strong line beyond about 20\,d after explosion (bottom left).
He\one\,10830\,\AA\ is strong at all times beyond 5\,d (bottom right).
\label{fig_line}
}
\end{figure*}

   By the end of our simulations, the only lines present in the optical and near-IR are He\one\,10830\,\AA\
   and the forbidden line transitions of Ca\two\ (7307\,\AA\ doublet) and Ti\two\ (four strong
   transitions in the near-IR). The distinguishing feature of these events is not the presence
   of  [Ca\two], which is a common property of most SNe, either core collapse or thermonuclear.
    [Ca\two] transitions are strong because they are the key coolants of SN ejecta, in particular
    when other species/lines are not present to compete (e.g., the 3d$^6$\,$^5$D--\.3d$^6$\,$^3$F
    transition of Fe\three\ at 4658.0\,\AA\ or the 3d$^7$\,$^4$F--\,3d$^7$ $^2$G
    transition of Co\three\ at 5888.5 in SNe Ia; H$\alpha$ in SNe II).
   The distinguishing feature of this model, and  the class of so-called Ca-rich transients, is
   the lack of other strong lines at nebular times. Other distinct features are the presence of He\one\ and
   Si\two\ lines in the photospheric phase, causing a hybrid classification as Type Ib and Ia.

   The morphology of individual lines is often affected by overlap and blanketing. Nonetheless,
    for strong lines, an identification is obvious and one can measure the velocity at maximum
    absorption to infer the typical expansion rate of the ejecta. We show a set of spectral montages
    focusing on He\one\,5875\,\AA, Si\two\,6355\,\AA, Ca\two\,7307\,\AA\ and He\one\,10830\,\AA\
    in Fig.~\ref{fig_line}, restricting the time range to epochs when the features
    are the strongest. All these lines suggest a typical expansion rate of 10000\,\kms, quite typical
    of Ca-rich transients studied here \citep{perets_etal_10,waldman_etal_11,kasliwal_etal_12}. Note that half
    the ejecta mass lies beyond 10000\,\kms\ in our model (Fig.~\ref{fig_comp_init}).

    Previous simulations of helium-shell detonations neglected non-thermal processes and failed
    to produce He\one\ lines  \citep{shen_etal_10,waldman_etal_11}.
    Here, despite the rather uniform He mass fraction, we find these lines are
    not always present.
    In the optical, they can be masked by Ti\two\ line blanketing, which is  strong in our model.
    The notable exception is the enduring He\one\,10830\,\AA\ line.
    Omitting this aspect, the simulations of \citet{waldman_etal_11} show comparable optical
    signatures during the photospheric phase.

\section{Calcium mass from nebular spectra}
\label{sect_ca}

   In the model spectrum at 55\,d, we measure a total luminosity in Ca\two\,7307\,\AA\ of 3.57$\times$10$^{39}$\,erg\,s$^{-1}$.
 The expression for the total luminosity in the line is
  \begin{equation}
  L(\hbox{Ca\,{\sc ii}}) = h \nu_{ul} N_{u} A_{ul} 
  \end{equation}
   where  $\nu_{ul}$  is the line frequency, $N_u$ is the upper level population and $A_{ul}$ is the radiative de-excitation rate.
   The critical densities for the Ca\two\ 7291.5--7323.9\,\AA\ lines are 4.6--4.0$\times$10$^6$\,cm$^{-3}$ at the
   temperature of 2550\,K where the lines form in our ejecta model at 55\,d. In the corresponding regions,
   the electron density is about 1-1.5$\times$10$^7$\,cm$^{-3}$, thus higher
   than the critical density. Hence, one can assume that the upper level of each transition is in LTE with respect
   to the ground state. The above expression then becomes:
    \begin{equation}
    L(\hbox{Ca\,{\sc ii}}) =  {h \nu_{ul} g_u A_{ul} \over g_l A_{\rm Ca}  m_{\rm u} }  \exp(- h \nu_{ul} / kT) M({\rm Ca})  
    \end{equation}
   where  $g_u$ and $g_l$ are the statistical weights of the lower and upper transitions,
   T is the gas temperature, M(Ca) is the total mass of calcium contributing to
   the line emission, $A_{Ca}$ is the atomic mass of Ca (which we take as 40; the \iso{44}Ti has not yet decayed to affect
   the Ca abundance), and $m_{\rm u}$ is the atomic mass unit.
   Adopting the value of 0.024\,\msun\ for the total calcium mass within 10000\,\kms\ (see Figs.~\ref{fig_comp_init} and \ref{fig_line}),
   we find a theoretical flux  of 4.26$\times$10$^{39}$\,erg\,s$^{-1}$, in good agreement with the line luminosity
   measured in the theoretical spectrum. Impressively, this Ca\two\ 7307\,\AA\ line alone radiates 80\,\%
   of the total SN luminosity at that time. The rest of the luminosity in our model comes out in the near-IR,
   primarily through forbidden transitions of Ti\two.

 \begin{figure}
\epsfig{file=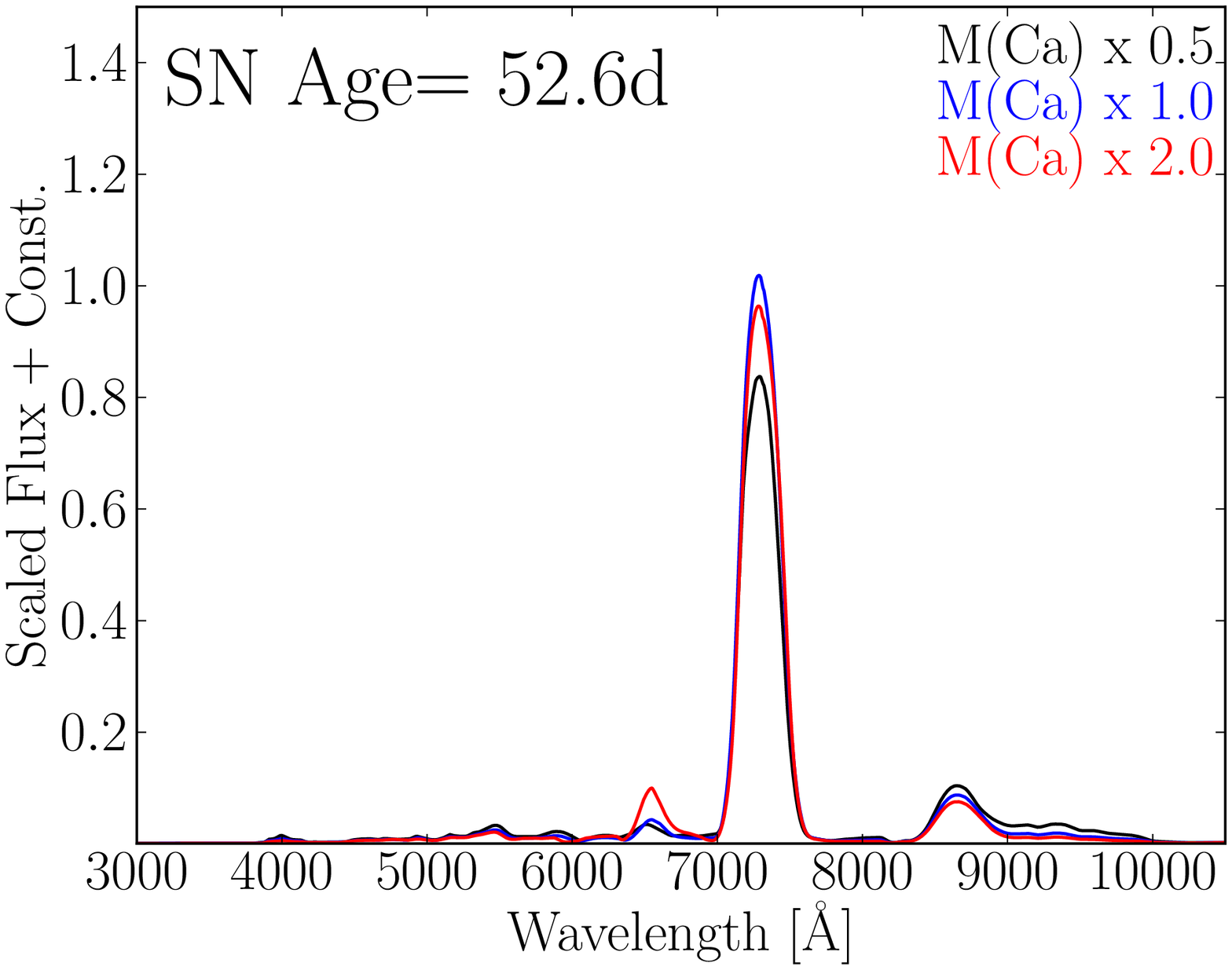,width=8.75cm}
\epsfig{file=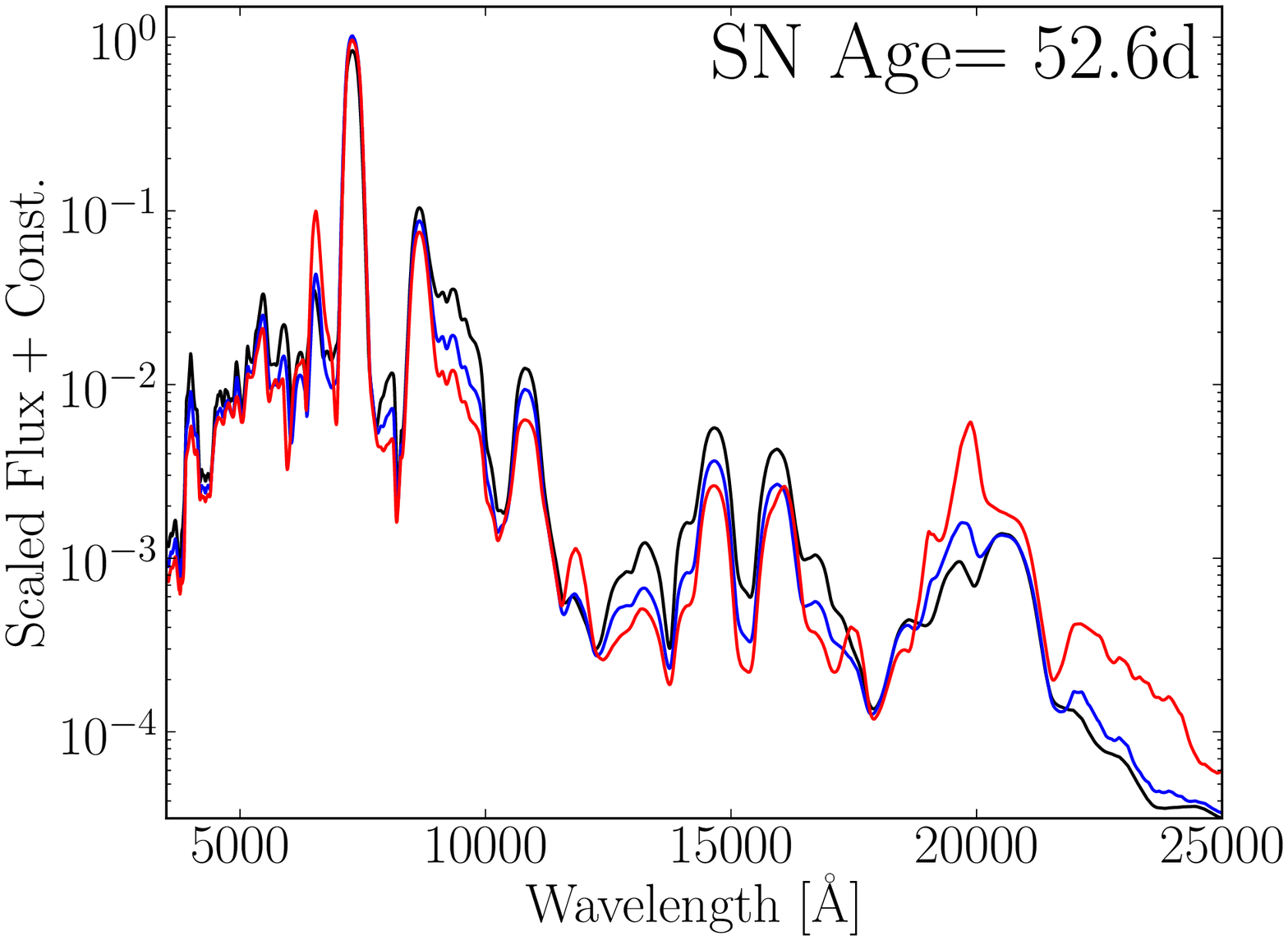,width=8.75cm}
\epsfig{file=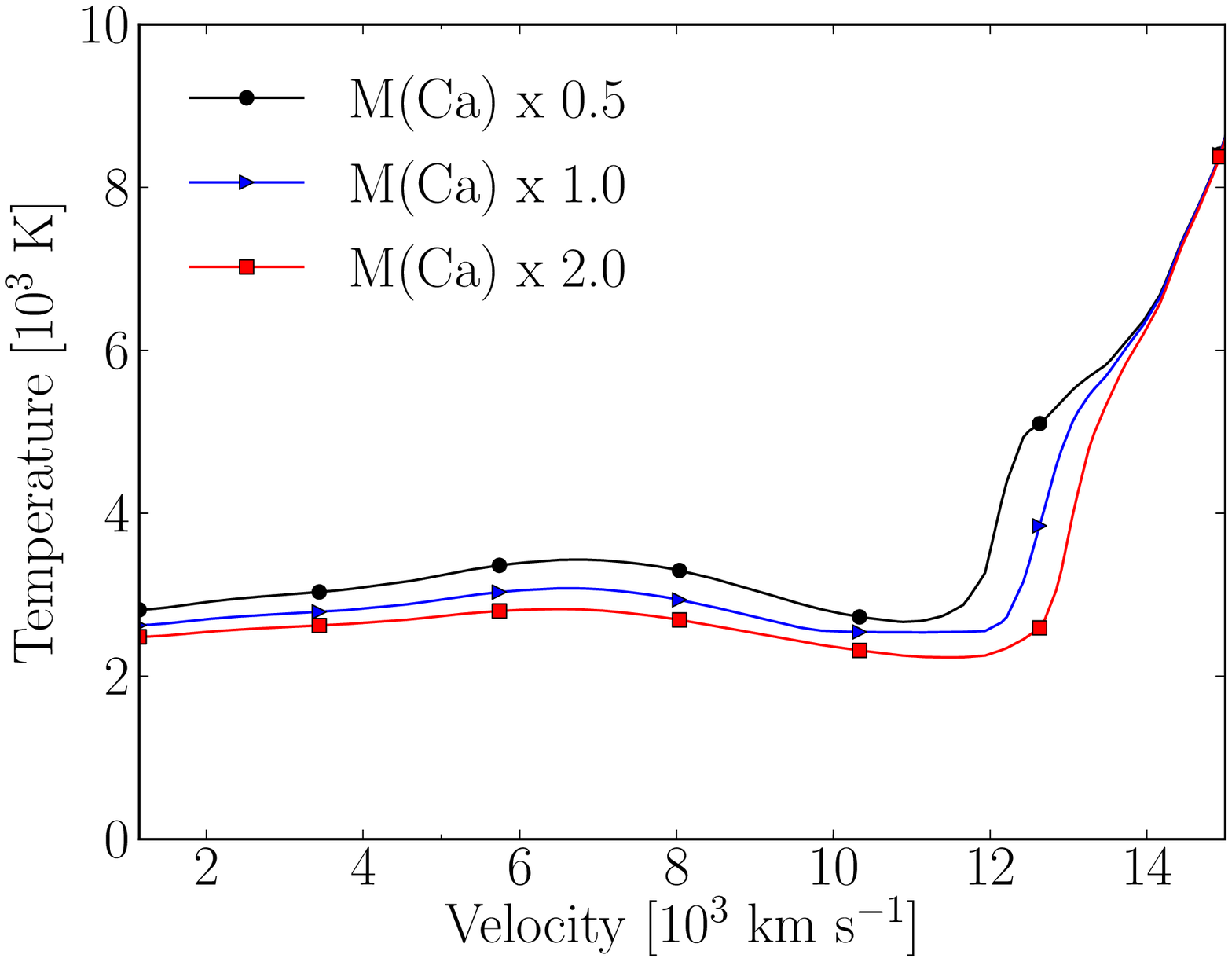, width=8.75cm}
\caption{
{\it Top:} Synthetic optical spectra for the model COp45HEp2 at 52.6\,d but with a scaling of the Ca abundance by
0.5, 1.0, and 2.0.
{\it Middle:} Same as left, but now for the optical and near-IR ranges, shown on a log scale.
{\it Bottom:} Corresponding temperature profiles. The same color coding applies to all three panels.
\label{fig_xca}
}
\end{figure}

   Because Ca\two\ dominates the cooling, and because we do not have a reliable independent means of determining
   the ejecta temperature (which appears in the exponential) from the observations, it is extremely difficult to determine
   an accurate calcium mass. In the present model
   the Ca\two\ line strength is set by the initial mass of  \iso{48}Cr (which sets the amount of $\gamma$-ray energy that
   can be absorbed) and the ejecta mass and structure (which determines the amount of energy absorbed) in the Ca
   emitting region. Unfortunately, Ca, Ti, and He all contribute to the $\gamma$-ray absorption in the inner region of
   the ejecta.

    We have tested the sensitivity of the emergent spectrum to the calcium abundance. In two additional simulations
   based on model COp45HEp2 at 52.6\,d,
   we have multiplied the Ca mass fraction throughout the ejecta by a factor of 0.5 and 2.0. In practice, we
   scale the \iso{40}Ca mass fraction and adjust the helium mass fraction (which is large at all depths) to preserve
   the normalization to unity (in practice, this also impacts the $\gamma$-ray trapping but we neglect this effect for
   this simple test). We find that the flux in the Ca\two\,7307\,\AA\ line varies by only $\sim$\,10\,\% between
   these simulations (Fig.~\ref{fig_xca}) and that this line remains the main coolant in each case --- a reduction in flux
   of the Ca\two\,7307\,\AA\ line is compensated by an increase in the near-IR flux of forbidden lines of Ti\two\
   (for a large enhancement in Ca mass fraction, the Ca ionization can shift and allow for cooling through
   Ca\,\one\ lines as well).
   Rather than changing the Ca\two\ line flux, varying the total Ca mass causes a change in temperature.
   This weak sensitivity of the Ca\two\,7307\,\AA\ line flux to the calcium abundance results from the fact that this
   line is the main coolant. As long as this holds, the corresponding Ca\,\two\ line flux reflects primarily the trapped decay
   energy (which must be radiated) and only more weakly the Ca abundance.

\section{Oxygen mass from nebular spectra}
\label{sect_oxy}

Our model has a very low oxygen mass, and consequently we do not predict any emission in
[O\one]\,6300--6364\,\AA. This doublet line seems to be present at nebular times in both PTF10iuv and SN\,2005E.

In order to get O emission two criteria need to be met. First, we need a significant amount of
O in a region where $\gamma$-rays are being absorbed. Second, the O emitting region cannot be too
contaminated by Ca. The [Ca\two] doublet at 7307\,\AA\ is a much more efficient coolant than [O\one]\,6300\,\AA\
(by nearly a factor of 1000 per atom in the high density limit),
and thus can limit the strength of [O\one]\,6300--6364\,\AA\ emission from a region in which both Ca$^+$ and
O$^{0+}$ co-exist \citep{Fanson_etal_89}.
While  oxygen features are readily identified in  photospheric phase spectra of
SN\,2008ha,  [O\one]\,6300\,\AA\ is absent from the spectrum at 62 days \citep{foley_etal_09}
in SN\,2008ha (but present in SN\,2005E).

An estimate of the oxygen mass, in the high density limit, can be made using
\begin{equation}
M[\rm O]=1.1 \times 10^8 f_{[\rm O \, \tiny I]} D^2_{\rm Mpc} \, \exp(2.78/T_4) 
\end{equation}
\noindent
where $f_{[\rm O \, \tiny I]}$ is the total flux in the [O\one] 6300--6364 doublet, $D_{\rm Mpc}$ is the distance
to the SN in Mpc, and $T_4$ is the temperature in 10$^4$\,K --- this expression is very similar
to the one given by \cite{Uomoto_86}.  In practice this is, however, not a particularly useful expression
since the derived O mass is very sensitive to temperature (just like for the Ca mass estimate; see preceding section).
For SN\,2005E, \citet{perets_etal_10} estimate an oxygen mass of 0.037\,\msun\ and use $T_4=$\,0.45.
If we were to adopt the temperature of 2550\,K in the [Ca\two] emitting zone of our model, this mass estimate
would be 110 times larger, which is unrealistic.

In reality, the emission regions of [O\one] and [Ca\two] are probably distinct.
The  O\one\ emission strength will be primarily
set by the non-thermal energy absorbed in the corresponding region.  A fundamental requirement of any model to
explain SN\,2005E is that the ejecta have a significant amount of O, and this O lies outside the [Ca\two]
emitting region. The uncertainty that surrounds the origin of these fast transients, in particular concerning the
explosion mechanism and the importance of multi-D effects, prevent a clear determination of the
spatial distribution of O and Ca, which in turn compromises the accuracy of the O abundance determination.

\begin{figure*}
\epsfig{file=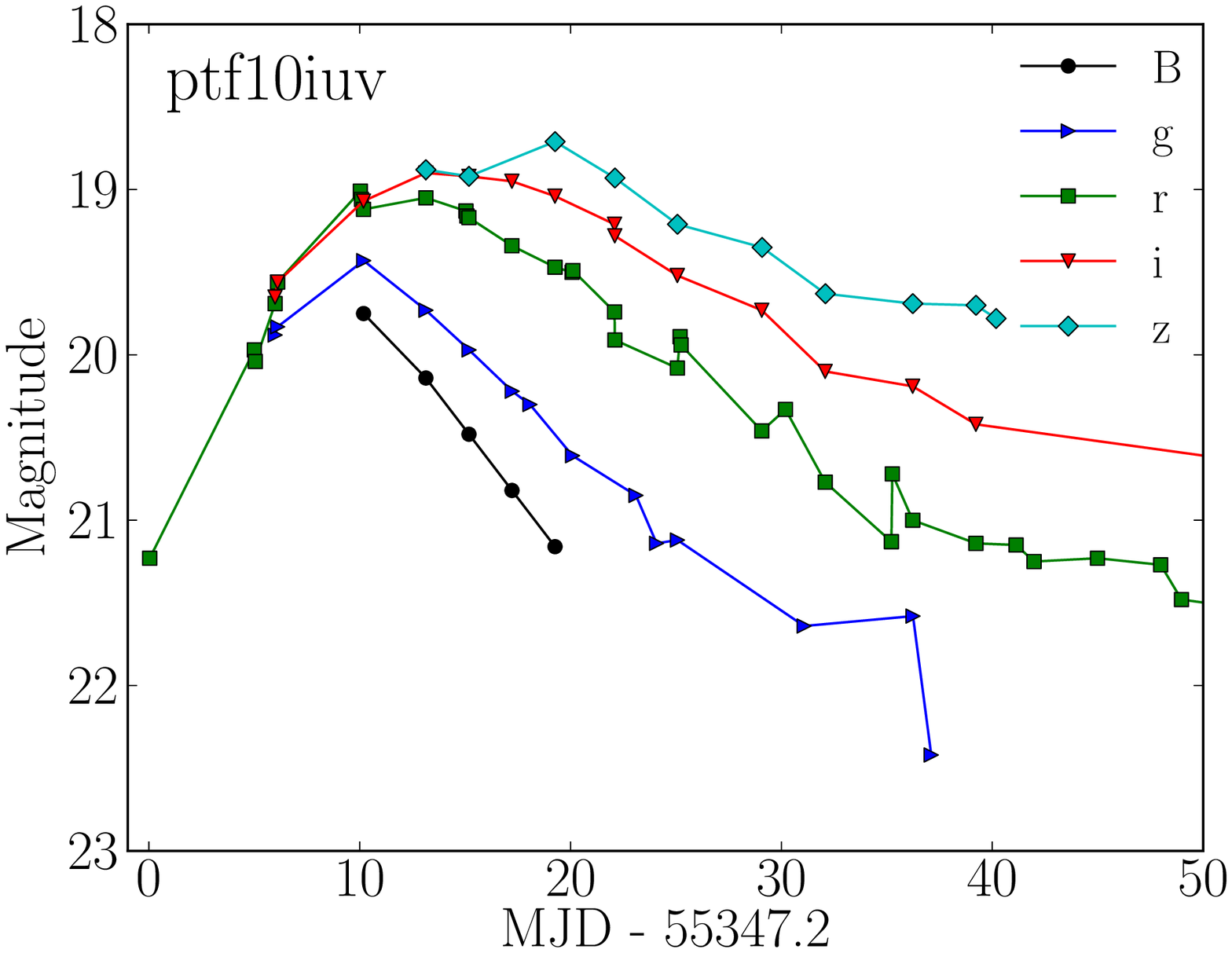,width=8.5cm}
\epsfig{file=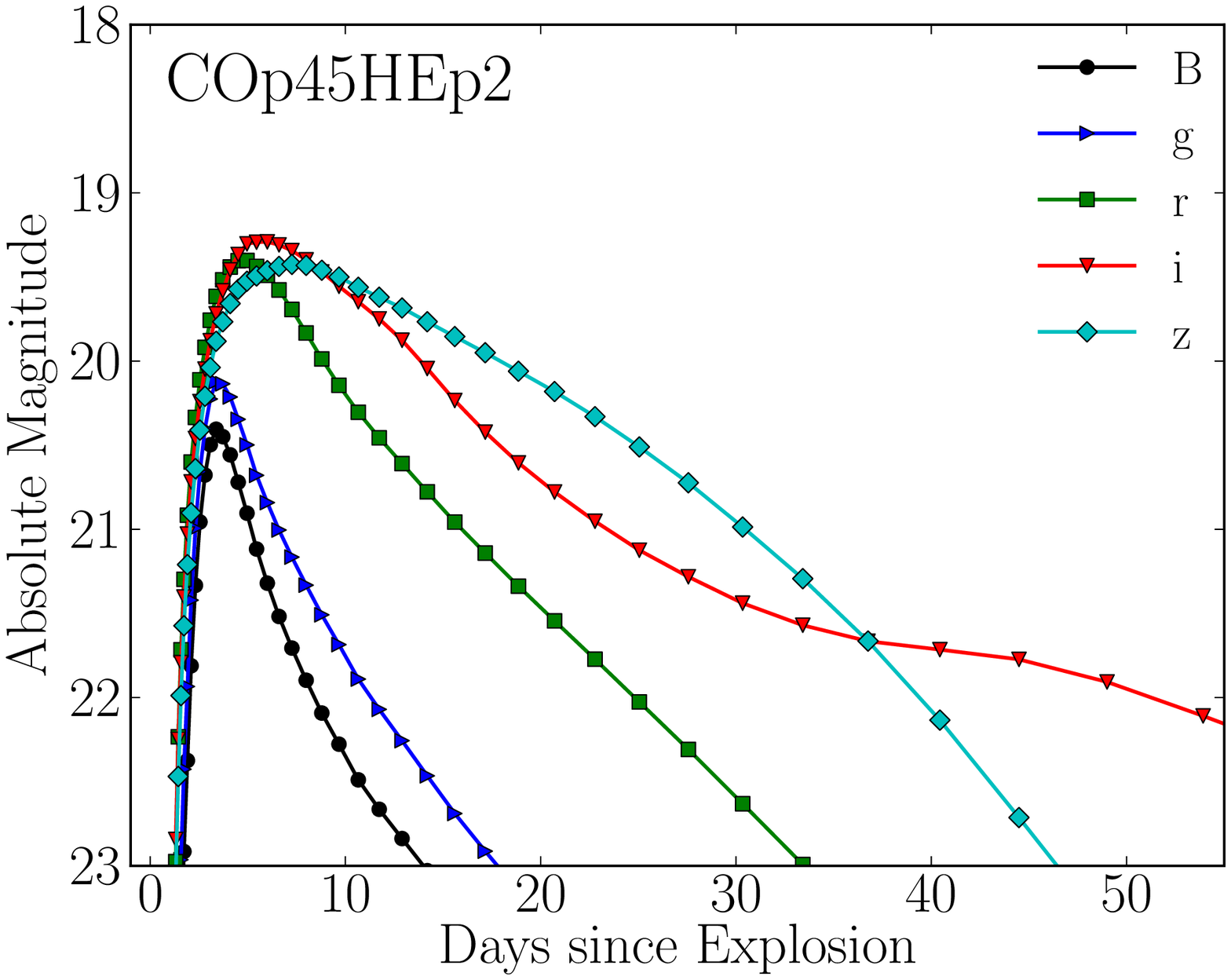,width=8.5cm}
\caption{{\it Left:} Multi-band light curves of PTF10iuv \citep{kasliwal_etal_12}.
{\it Right:} Synthetic photometry for model COp45HE2 but scaled to match the
distance and redshift to PTF10iuv. While model COp45HEp2 exhibits a rapid rise like PTF10iuv
it is too faint, and fades more rapidly. This discrepancy suggests the model ejecta mass is too small.
[See text for details.]
\label{fig_lc_Bgriz}
}
\end{figure*}

\section{Comparison to Observations and conclusions}
\label{sect_comp}

  Numerical simulations of He-shell detonations suggest a wide range of ejecta yields and
  masses \citep{fink_etal_07,shen_etal_10,waldman_etal_11}. The outcome is function of both the WD mass
  and the He-shell mass, and thus theoretical models can yield a
   large diversity of transient light curves and spectra.

  Observationally, Ca-rich transients show a common set of properties \citep{kasliwal_etal_12},
  in particular the faint peak luminosity, a fast rise to maximum, a narrow light curve width, fast expansion,
  and a preponderance of Ca\two\ lines in nebular-phase optical spectra. All of these properties
  are a suitable description of .Ia SN models.

   The multi-band light curves of PTF10iuv show a moderate resemblance  to those of model COp45HE2
   (Fig.~\ref{fig_lc_Bgriz}). For example, both observations and model exhibit a time to maximum brightness
   that increases for redder filters. The model peak brightness is within 0.5-1\,mag of the observed value, depending
   on the filter. Important disagreements are the longer rise to peak, the broader and the more slowly declining light curves
   (in all bands) observed in PTF10iuv compared to the model.  These mismatches are also present if we compare to
   SN\,2005E \citep{perets_etal_10}.

\begin{figure*}
\epsfig{file=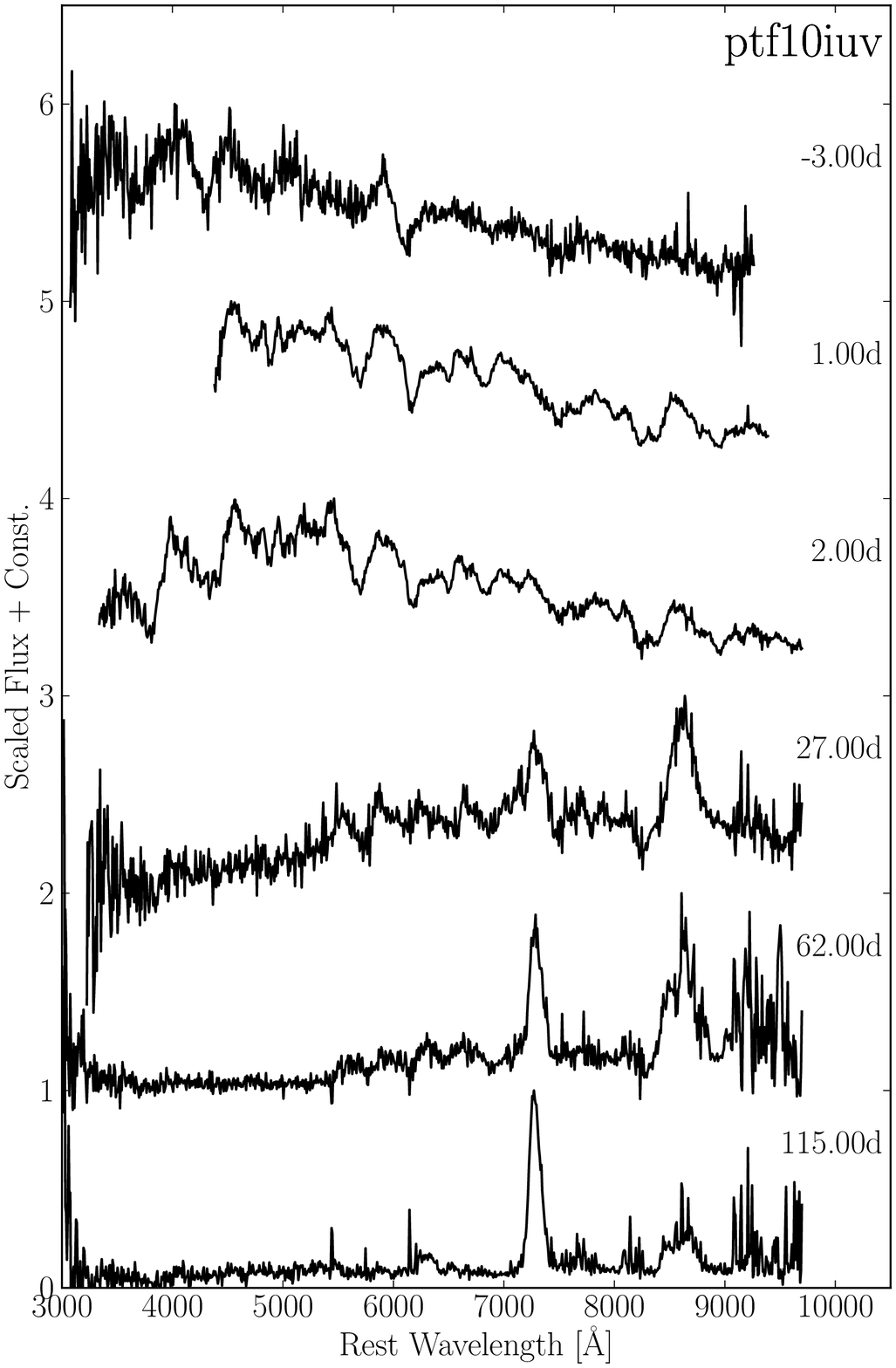,width=5.5cm}
\epsfig{file=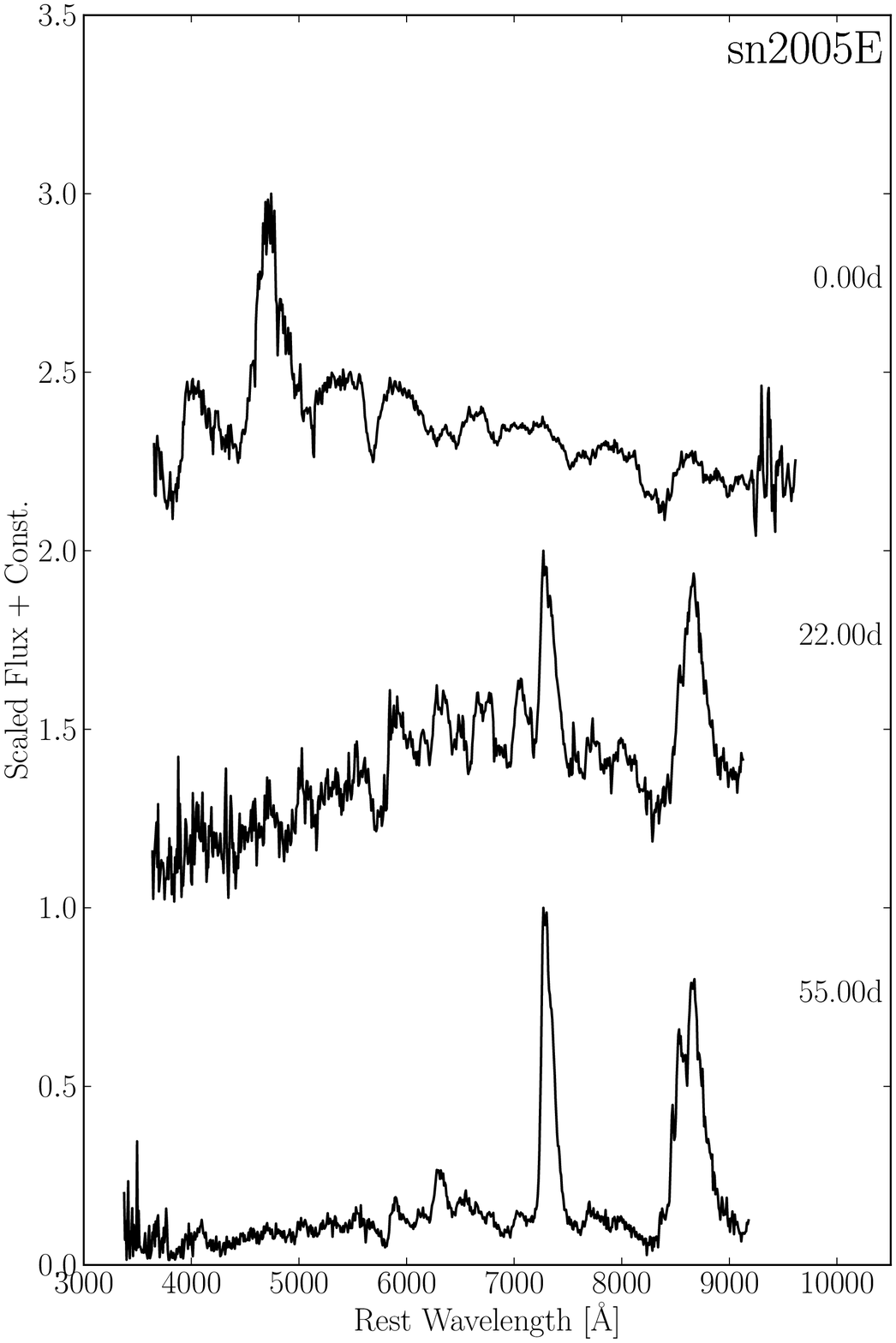,width=5.5cm}
\epsfig{file=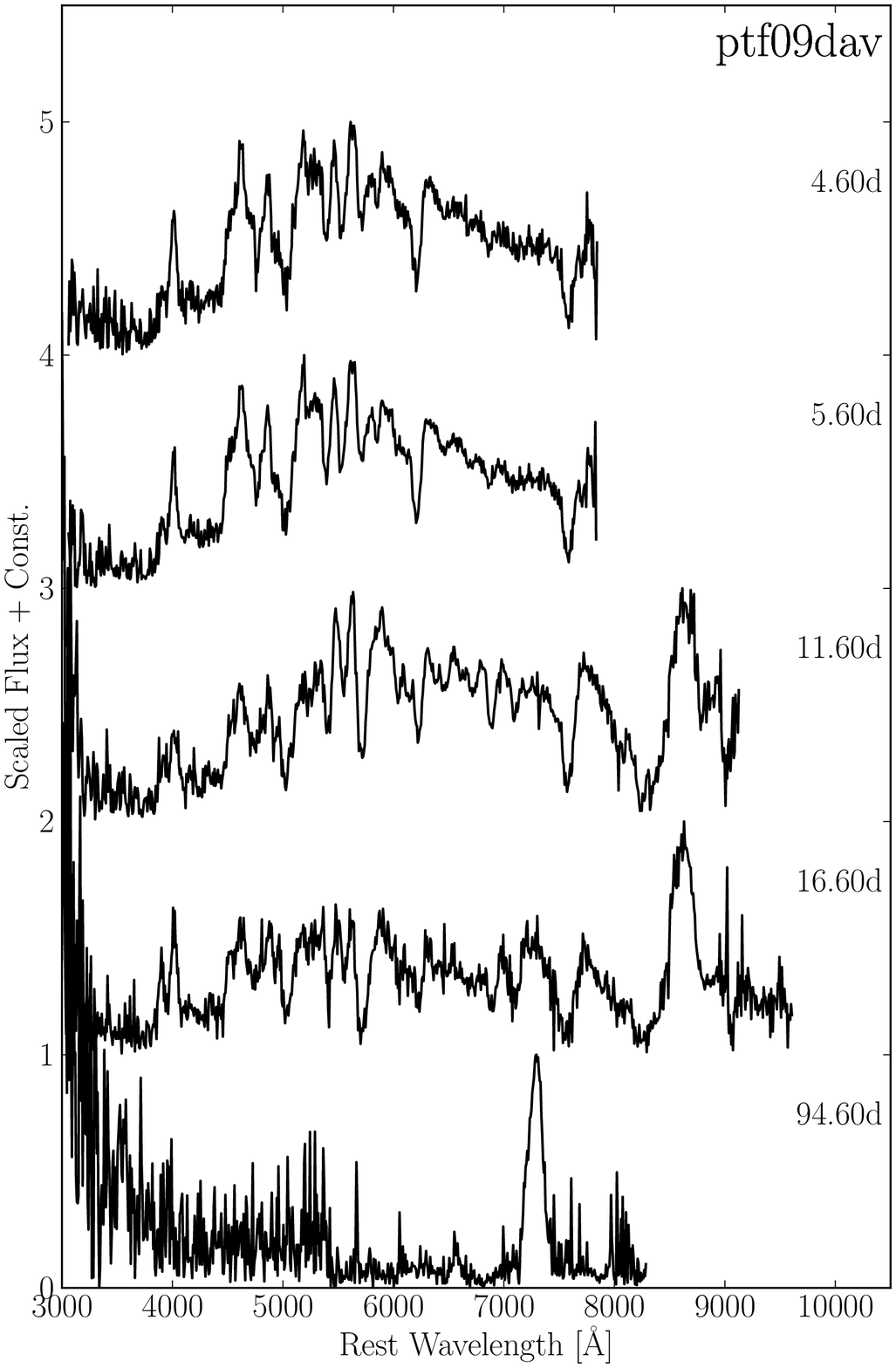,width=5.5cm}
\epsfig{file=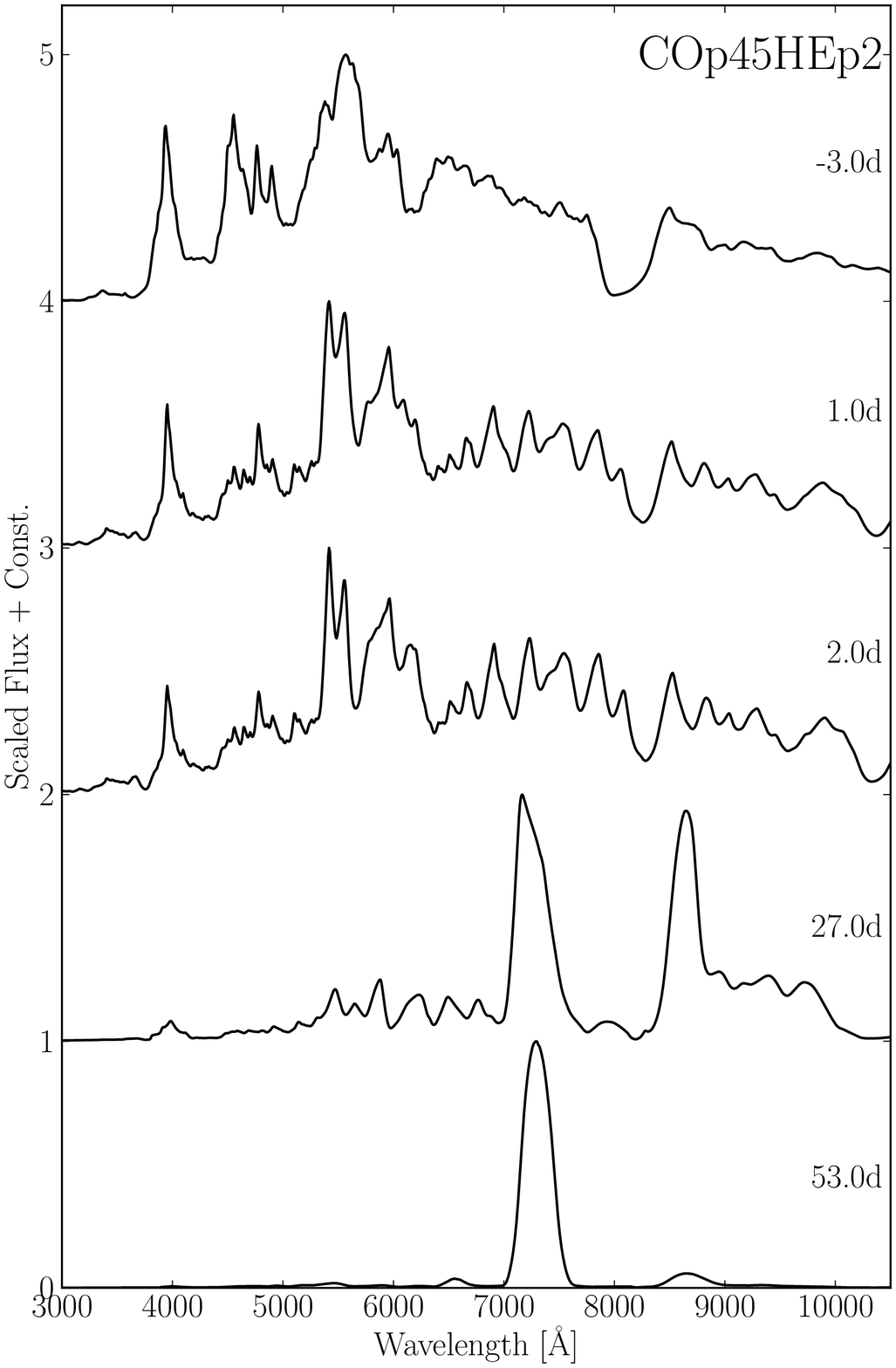,width=5.5cm}
\epsfig{file=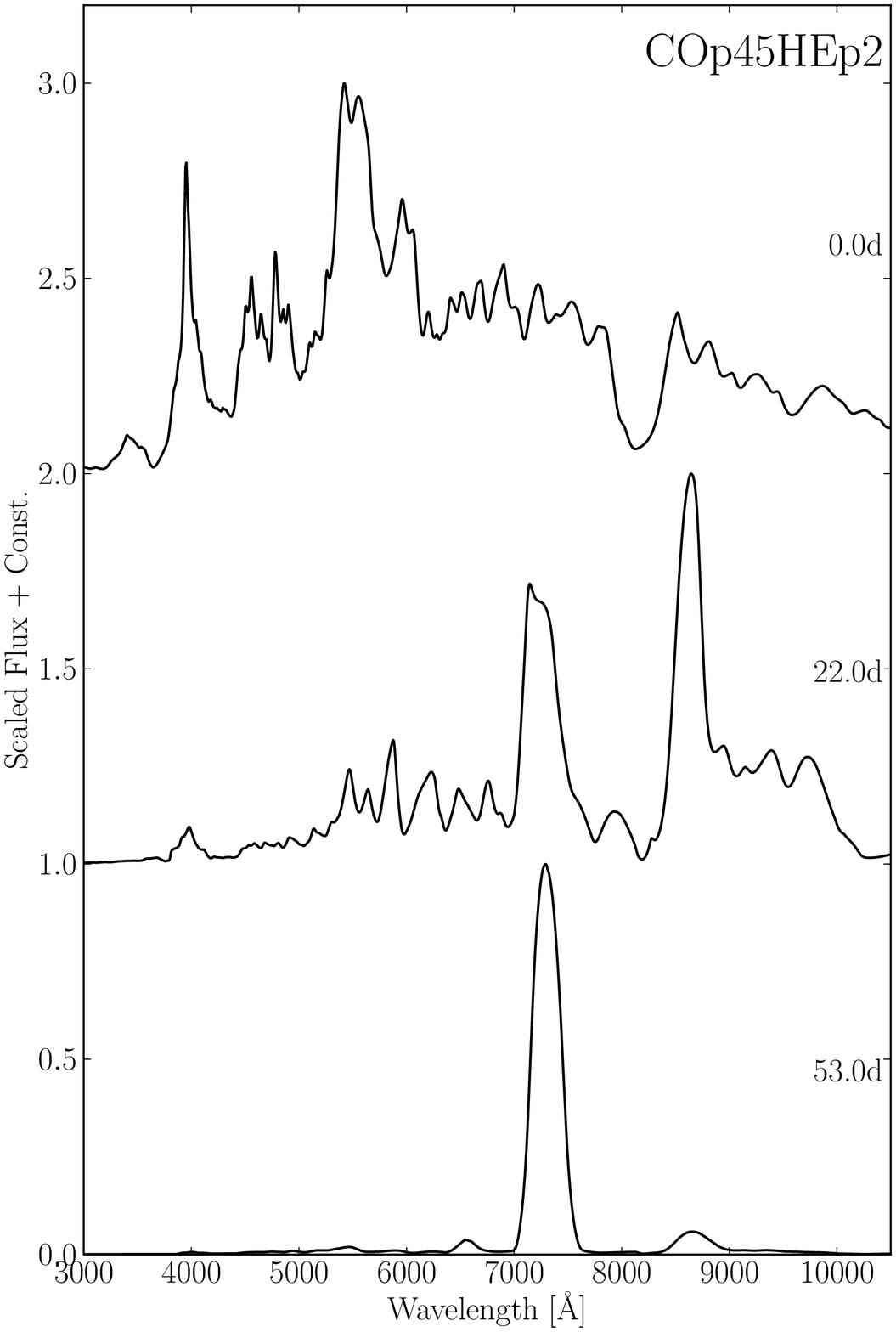,width=5.5cm}
\epsfig{file=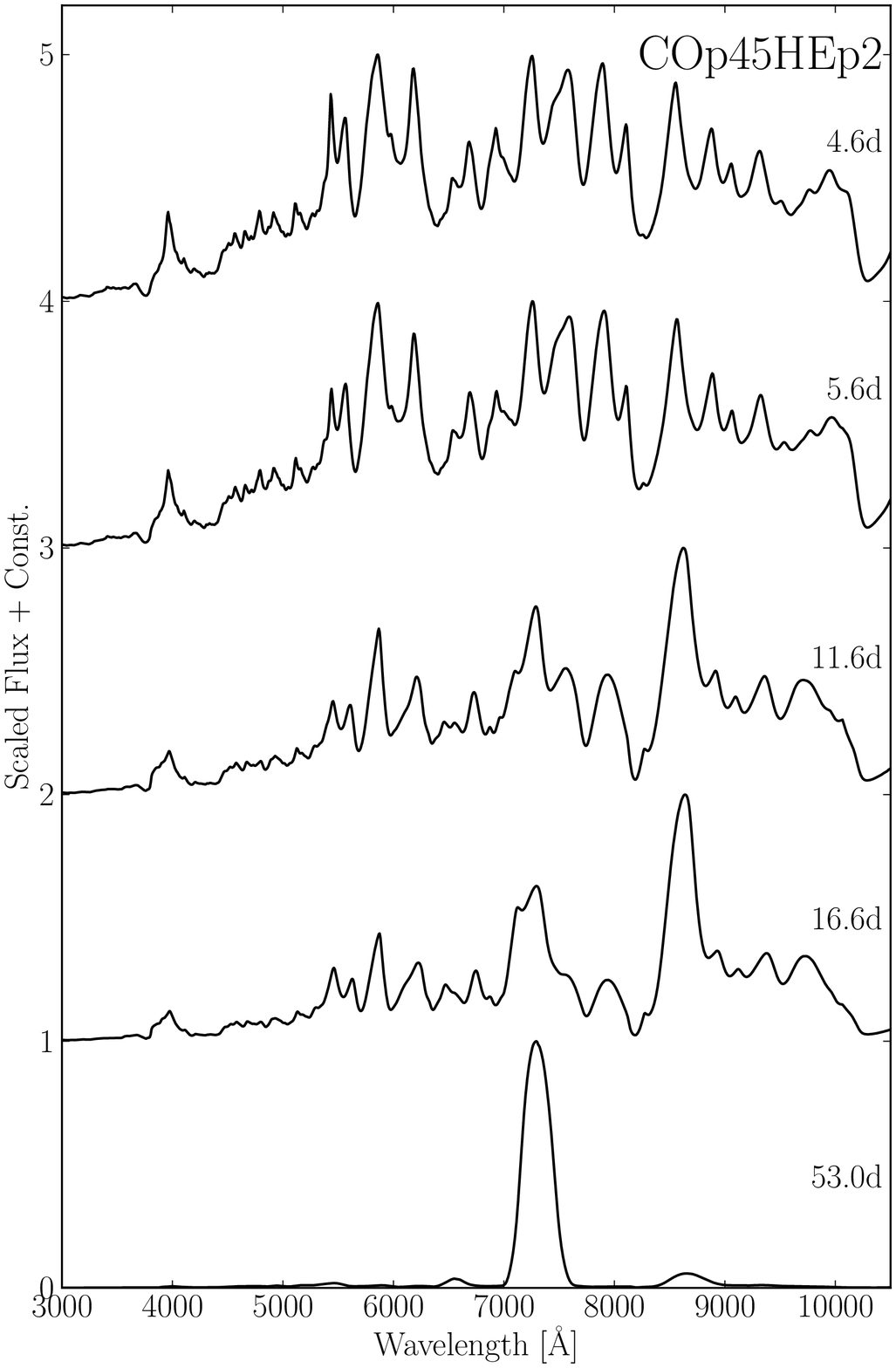,width=5.5cm}
\caption{{\it Top row:} Montage of rest-frame spectra for PTF10iuv \citep{kasliwal_etal_12},
SN\,2005E \citep{perets_etal_10}, and PTF09dav \citep{sullivan_etal_11}.
{\it Bottom row:} Montage of spectra for model COp45HEp2 shown at epochs that
correspond to those of observations, column by column.
The plots are offset vertically, and for each plot the associated zero is set at 3000\,\AA.
[See text for details.]
\label{spec_obs_mod}
}
\end{figure*}

   In Fig.~\ref{spec_obs_mod}, we compare model spectra  with observations (at comparable epochs with
   respect to maximum brightness) for PTF10iuv, SN\,2005E, and PTF09dav.
   Observed pre-peak spectra tend to be bluer than in the model. Our model is red even at peak, which is expected
   given the high metal abundance from the explosion.\footnote{To test the influence of Ti,
   we artificially scaled down its abundance in the formal solution of the radiative transfer, i.e., to calculate
   the emergent spectrum. Scaling it down by a factor of 0.1 or 0.01 in the model at 5.9\,d
   makes the spectrum increasingly bluer (in this order, the $V-I$ magnitude varies from 0.74\,mag to 0.55
   and 0.30\,mag) , and the Ti\two\ lines in the red weaker. }
   However, this makes it hard to compare to early-time observations because the temperature difference
   is tied to an ionization offset and thus different ions/lines influence the spectra.
   He\one\ lines, which stem from non-thermal processes, are predicted by the model,
   but the observations show their presence for much longer.
   The model predicts strong blanketing due to Ti\two, stronger than observed, which suggests Ca-rich transients
   in Nature may have a lower mass fraction of metals than in our model.
   In PTF09dav, Sc\two\ lines are clearly visible \citep{sullivan_etal_11} but are weak or absent in our model.
   The spectral differences at early time between PTF10iuv and PTF09dav suggests that there
   is diversity in Nature amongst Ca-rich transients, likely stemming from variations in burning 
   yields.\footnote{Our 
   model results are in closer agreement with the observations of OGLE-2013-SN-079, which show
   red colors and a spectrum apparently influenced by strong Ti\two\ line blanketing \citep{inserra_etal_14}.
   We do not discuss this object in detail here because the paper by Inserra et al. was submitted while this work
   was under review.}

   Models and observations agree well at nebular times in that they both show a dominance of Ca\two\ lines.
    We note a few discrepancies between synthetic and observed spectra at nebular times.
   The Ca\two\ line at 7307\,\AA\ is broader in the model than observed.
   The half-width at zero flux is $\sim$\,5000\,\kms\ in PTF10iuv
   compared to $\sim$\,10000\,\kms\ in the model. The Ca\two\ forbidden-line doublet (Ca\two\ triplet at 8500\,\AA)
   also appears (disappears) much earlier  than in the observations.
   Furthermore, our model has a very low O mass,  and consequently does not predict the [O\one]\,6300--6364\,\AA\ that
   seems to be present at nebular times in PTF10iuv and SN\,2005E.

   Together with the mismatches in early-time color, light curve width, and post-peak decline rate, these spectral
   differences suggest that the ejecta mass of these Ca-rich transients is most likely larger than 0.2\,\msun\ and the
   helium shell more extended initially. The fainter peak of our model could be cured by a larger \iso{48}Cr.
   \citet{waldman_etal_11} argue that by enhancing the Ti abundance, the post-peak brightness could be brought
   into agreement with the observations of SN\,2005E. However, this solution is unlikely because
   the Ti\two\ lines are already stronger in the model than in the observations.
   A lower expansion rate and thus a lower explosion energy would also help reconcile some of
  the mismatches, but our model is already too faint compared to PTF10iuv. So, it seems
  a higher ejecta mass is really needed to get the broader light curve, yielding a slower expansion rate that also agrees
  better with the relatively narrow line profiles observed.

\begin{figure}
\epsfig{file=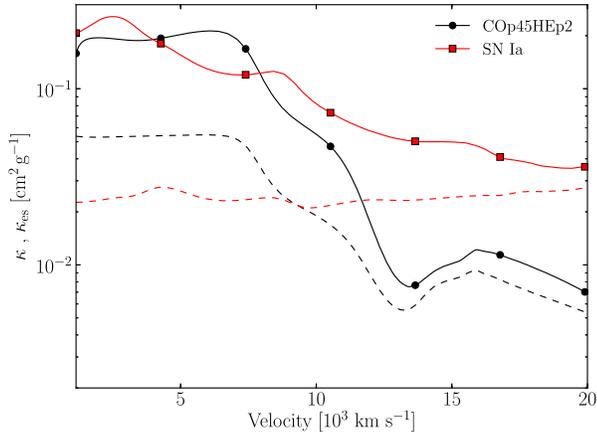,width=8.5cm}
\caption{Comparison of the total opacity (i.e., mass absorption coefficient $\kappa$)
and the electron scattering opacity ($\kappa_{\rm es}$; dashed line) between
the He-detonation model COp45HEp2 and the standard-luminosity SN Ia model DDC10 \citep{dessart_etal_14c}
at the time of bolometric maximum.
The difference at large velocities stems from the low ionization in the COp45HEp2 model.
The similarity at lower velocities is a combination of several effects.
Line opacity from IGEs in a SN Ia ejecta is large but this is offset by the larger mean atomic weight.
These regions are also ionized and optically thick at maximum light in both models so electron
scattering matters, and its contribution is larger in the He-detonation model.
\label{fig_kappa}
}
\end{figure}

   Most estimates for the ejecta mass of Ca-rich transients typically assume that the opacity is the same as in standard SNe Ia.
   However, the opacity of standard SNe Ia is dominated by IGEs, and to a lesser extent, IMEs.
   Conversely, our He shell detonation model has a large He abundance, and hence we
   would expect a significantly lower opacity (Fig.~\ref{fig_kappa}; see also bottom left panel of Fig.~\ref{fig_prop_evol}).
   The ejecta mass inferred from the light curve
   scales as $1/\kappa$ (the light curve morphology is sensitive to optical depth, which goes as $M \kappa$; \citealt{arnett_82}),
   so ejecta masses based on scaling from normal SNe Ia will tend to underestimate the true ejecta mass.

  A crucial question, which should be obtainable from accurate light curves and spectral modeling, is
  what are the radioactive isotopes that heat the SN ejecta and produce the bright display.
  One needs to determine the relative contributions from, for example,
  the decay chain associated with \iso{56}Ni  (which powers standard SNe Ia light curves),
  and the decay chain associated with \iso{48}Cr.

 One interesting possibility is that these
Ca-rich transients do not stem from a detonation of an helium-accreting CO WD, but instead from the
merger of a CO WD and a He WD (see the merger simulations, e.g., of \citealt{pakmor_etal_13}).
The lighter He WD is less dense and gets completely disrupted
during the coalescence. When He eventually comes to rest at the surface of the CO remnant, a detonation
is somehow born (just like in the He-shell model) giving rise to an explosion in an extended and much
more massive He shell.
The detonation inside an extended shell would likely cause some exchange
from kinetic energy into internal energy, as in pulsational-delayed-detonation models of SNe Ia
or in WD explosions within a buffer of mass (see, e.g., \citealt{fryer_etal_10}),
and contribute excess luminosity and bluer colors at early times \citep{hoeflich_khokhlov_96,dessart_etal_14b}.
The greater mass of the He shell would cause a more efficient trapping of $\gamma$-rays and
thus a more sustained  brightness past peak, in better agreement with observations of Ca-rich transients.

\section*{Acknowledgments}

We acknowledge discussions with Carles Badenes, Ryan Foley, Mansi Kasliwal, and Roni Waldman.
We thank Roni Waldman and Eli Livne for providing model COp45HEp2, and the observers for making their data
public through {\sc wiserep}.
LD acknowledges financial support from the European Community through an
International Re-integration Grant, under grant number PIRG04-GA-2008-239184,
and from ``Agence Nationale de la Recherche" grant ANR-2011-Blanc-SIMI-5-6-007-01.
DJH acknowledges support from STScI theory grant HST-AR-12640.01, and NASA theory grant NNX14AB41G.
This work was supported in part by the National Science Foundation under Grant No. PHYS-1066293
and the hospitality of the Aspen Center for Physics.
This work was granted access to the HPC resources of CINES under the allocations c2013046608
and c2014046608 made by GENCI (Grand Equipement National de Calcul Intensif).

\label{lastpage}

\end{document}